\documentclass[conference]{IEEEtran}
\makeatletter
\def\ps@headings{%
\def\@oddhead{\mbox{}\scriptsize\rightmark \hfil \thepage}%
\def\@evenhead{\scriptsize\thepage \hfil \leftmark\mbox{}}%
\def\@oddfoot{}%
\def\@evenfoot{}}
\makeatother
\usepackage[cmex10]{amsmath}
\usepackage{array}
\usepackage{url}
\usepackage{makecell}
\usepackage{graphicx}
\usepackage{subfig}
\usepackage{amssymb}
\usepackage{multirow}
\usepackage{multicol}
\graphicspath{{./figures/}}

\begin{document}

%
\title{LogMaster: Mining Event Correlations in Logs of Large-scale
Cluster Systems}



%
\author{\IEEEauthorblockN{Rui Ren\IEEEauthorrefmark{1},
Xiaoyu Fu\IEEEauthorrefmark{1},
Jianfeng Zhan\IEEEauthorrefmark{1}, \IEEEauthorrefmark{3},
Wei Zhou\IEEEauthorrefmark{1}}
\IEEEauthorblockA{\IEEEauthorrefmark{1}Institute of Computing Technology, Chinese Academy of Sciences}
\IEEEauthorblockA{\IEEEauthorrefmark{3} Corresponding author: zhanjianfeng@ict.ac.cn}
}


\maketitle

\begin{abstract}


This paper presents a methodology and a system, named \emph{\textbf{LogMaster}}, for mining correlations of events that have multiple attributions, i.e., node ID, application ID, event type, and event severity, in logs of large-scale cluster systems.
Different from traditional transactional data, e.g., supermarket purchases, system logs have their unique characteristic, and hence we propose several innovative approaches to mine their correlations.
We present a  simple metrics to measure correlations of events that may happen interleavedly.
On the basis of the measurement of correlations, we propose two approaches to mine event correlations;
meanwhile, we  propose an innovative abstraction---event correlation graphs (ECGs) to represent event correlations, and present an ECGs-based algorithm for predicting events.
For two system logs of a production Hadoop-based cloud computing system at Research Institution of China Mobile and a production HPC cluster system at Los Alamos National Lab (LANL), we evaluate our approaches in three scenarios: (a) predicting all events on the basis of both failure and non-failure events; (b) predicting only failure events on the basis of both failure and non-failure events; (c)  predicting failure events after removing non-failure events.


\end{abstract}






\section{Introduction} \label{introduction}

Cluster systems are common platforms for both high performance computing and cloud computing. 
As the scales of cluster systems increase, failures become normal \cite{Liang_DSN05}.
It has been long recognized that (failure) events are correlated, not independent. In an early year as 1992, Tang {\em et al.} \cite{Tang_TOC92} concluded that the impact of correlated failures on dependability is significant.
Though we can not infer causalities among different events without invasive approaches like request tracing \cite{Zhang_DSN09},  we indeed can find out correlations among different events through a data mining approach. This paper focuses on mining  \emph{recurring event sequences with timing orders that have correlations}, which we call \emph{event rules}. As an intuition, if there is an event rule that identifies the correlation of warning events and a fatal event,
the occurrence of warning events indicates that a failure may happen in a near future, so mining event rules is the basis for predicting failures.


Considerable work has been done in mining frequent itemset in transactional data, e.g., supermarket purchases, \cite{Agrawal_VLDB94} \cite{Agrawal_SIGMODE93}. However, system logs are different from transactional data in four aspects. First, for event data, the timing order plays an important role. Moreover, a predicted event sequence without the timing constraints provides little information for a failure prediction, so we have to consider time among event occurrences rather than just their occurrence \cite{Hellerstein_IBMSystemJournal02}. Second, logs are temporal. Only on condition that events fall within a time window, we can consider those events may have correlations. Third, a failure event has many important attributions, i.e., node ID, application ID, event type, and event severity. For a failure prediction, those attributions are ingredient. For example, if you predict a failure without node information, an administrator can not take an appropriate action. Last, different events may happen interleavedly, and hence it is difficult to define a metrics for event correlation. For example, it is difficult
to define the event correlation between two events $A$ and $B$ in an event sequence $BACBBA$, since $A$ and $B$ happen interleavedly.

On the basis of Apriori-like algorithms \cite{Agrawal_VLDB94} \cite{Agrawal_SIGMODE93}, several previous efforts propose new approaches for frequent itemset, event bursts, periodical events, mutual-dependent events \cite{Hellerstein_IBMSystemJournal02}, frequent episodes \cite{Mannila_DKD97}, sequential patterns\cite{Agrawal_ICDE95} or closed sequential patterns \cite{Tzvetkov_ICDM03} \cite{Yan_SDM03}, however mining multi-attribution event rules in system logs of large-scale cluster systems has its unique requirements as mentioned above.  For example, in \cite{Yan_SDM03}, instead of mining the complete set of frequent subsequences, Yan {\em et al.} mine frequent closed subsequences only, i.e., those containing no super-sequence with the same occurrence frequency, which is difficult to directly apply in mining event rules for the purpose of predicting failures. Some work proposed Apriori-like algorithms in predicting failures \cite{Gujrati_ICPP07} without providing details (multi-attribution) or rare events \cite{Sahoo_SIGKDD03} \cite{Vilalta_IDDM02}, which are limited to the specified target events.

Our effort in this paper focuses on mining event rules in system logs. Taking into account the unique characteristic of logs, we propose a simple metrics to measure event correlations in a sliding time windows.
Different from Apriori that generates item set candidates  of a length $k$ from all item sets of a length $k-1$ \cite{Agrawal_VLDB94} \cite{Agrawal_SIGMODE93},
we proposed a simplified algorithm that generates a $n-ary$ event rule candidate if and only if its two $(n-1)-ary$ adjacent subsets are frequent,
and hence we significantly decrease the time complexity.
We validate our approaches on the logs of a 260-node Hadoop cluster system at Research Institution of China Mobile and a production HPC cluster system---Machine 20 of 256 nodes at Los Alamos National Lab, which we call the Hadoop logs and the HPC logs, respectively. For predicting events in the Hadoop logs and the HPC logs, the precision rates are high as 78.20\%, 81.19\%, respectively. We also evaluate our approaches in three scenarios: (a) predicting all events on the basis of both failure and non-failure events; (b) predicting only failure events on the basis of both failure and non-failure events; (c)  predicting failure events after removing non-failure events.

Our contributions are four-fold. First, we propose a simple metrics to measure event correlations. Second, on the basis of the measurement metrics, we propose two approaches (\emph{Apriori-LES} and \emph{Apriori-semiLES}) to mining event correlations. Third, we design an innovative abstraction--\emph{events correlation graphs (ECGs)}, to represent event rules, and present an ECGs-based algorithm for event prediction. Fourth,  for the first time, we compare the breakdown of events of different types and events rules in two typical cluster systems for cloud and HPC, respectively.

The rest of this paper is organized as follows. Section~\ref{sec:Basic} explains basic concepts. Section~\ref{sec£ºsystem_design} presents LogMaster design. Section~\ref{sec:implementation} gives out LogMaster implementation. Experiment results and evaluations on the Hadoop logs and the HPC logs are summarized in Section~\ref{sec:evaluation}. In Section~\ref{sec:related_work}, we describe the related work. We draw a conclusion and discuss the future work in Section~\ref{sec:conclusion}.


\section {Background, and Basic Concepts}\label{sec:Basic}

\subsection{Background of two real system logs} \label{sec:background}



The 260-node Hadoop cluster system is used to run MapReduce-like cloud applications, including 10 management nodes, which are used to analyze logs or manage system, and 250 data nodes, which are used to run Hadoop applications. We collect the Hadoop logs by using /dev/error, /var/log, IBM Tivoli, HP openview, and NetLogger, then store them on the management nodes. So these logs include system service and kernel logs, such as crond, mountd, rpc.statd, sshd, syslogd, xinetd, and so on.  The HPC logs are available from \url{(http://institutes.lanl.gov/data/fdata/)}. TABLE~\ref{log_summary} give the summary of system logs from the Hadoop and HPC cluster systems.

\begin{table}[hbtp]
\renewcommand{\arraystretch}{1.3}
\centering
\begin{tabular}{|l|p{0.6cm}|p{1.0cm}|p{1.0cm}|p{0.6cm}|p{1.5cm}|}
  \hline
  \itshape Log name& \itshape Days &\itshape Start Date&\itshape End Date &\itshape Log Size &\itshape No. of Records  \\ \hline
  \itshape Hadoop& \itshape 67 &\itshape 2008-10-26 &\itshape 2008-12-31 &\itshape 130 MB &\itshape 977858 \\ \hline
  \itshape HPC& \itshape 1005 &\itshape 2003-07-31 &\itshape 2006-04-40 &\itshape 31.5 MB
   &\itshape433490 \\ \hline
\end{tabular}
\caption{The summary of logs}
\label{log_summary}
\end{table}


We use nine-tuples \emph{(timestamp, log ID, node ID, event ID, severity degree, event type, application name, process ID, user ID)} to describe each event.   For an upcoming event, if a new 2-tuple ($severity \: degree$, $event \: type$) is reported, our system will generate and assign a
new $event \: ID$ associated with this event.
  Similarly, if a new 4-tuple $(node \: ID, event \: ID, application \: name, process \: ID)$ is reported, we will create and assign a new $log \: ID$ associated with this event, and hence a $log \: ID$ will contain rich information, including severity degree, event  type, node ID, application  name, and process ID.

\begin{table}[hbtp]
\renewcommand{\arraystretch}{1.3}
\centering
\begin{tabular}{|l|p{5.6cm}|}
  \hline
  \itshape element item& \itshape description  \\ \hline
  \itshape timestamp &The occurrence time associated with the event \\ \hline
  \itshape severity degree  &Include five levels: INFO, WARNING, ERROR, FAILURE, FATAL \\ \hline
  \itshape event type  &Include HARDWARE, SYSTEM, APPLICATION, FILESYSTEM, and NETWORK \\ \hline
  \itshape event ID &An event ID is a mapping function of a 2-tuple (severity degree, event type). \\ \hline
  \itshape node ID  &The location of the event \\ \hline
  \itshape application name	& The name of the application that associates with the event \\ \hline
  \itshape process ID	&The ID of the process that associated with the event \\ \hline
  \itshape log ID  & A log ID is a mapping function of a 4-tuple (node ID, event ID, application name, process ID). \\ \hline
  \itshape user	&The user that associated the event \\ \hline
\end{tabular}
\caption{The descriptions about the elements of nine-tuples}
\label{describtion_nine_tuples}
\end{table}

\subsection{ Basic concepts} \label{definition}

\newtheorem{LES}{Definition}
\begin{LES}
\emph{An n-ary log ID sequence (LES)}: an $n-ary$ LES is a sequences of $n$ events \emph{that have different log IDs} in a chronological order. An $n-ary$ LES is composed of an $(n-1)-ary$ LES ($ n\in N+,n>1 $) and an event of log ID $X$ ($X \not\in (n-1)-ary \quad ES$ ) with the presumed timing constraint that an event of log ID $X$ follows after the $(n-1)-ary$ LES. We call the $(n-1)-ary$ LES is the \emph{preceding events} and the event of log ID $X$ is the \emph{posterior event}.
\end{LES}

For example, for a $3-ary$ LES $(A, B, C)$, $(A, B)$ are the preceding events, while $C$ is the posterior event. In this paper, we simply use an event $X$ instead of an event of log ID $X$.

\begin{LES}
\emph{A subset of  LES }: If the event elements of an $m-ary$  LES  are a subset of the event elements of an $n-ary$  LES  $(m<n)$, meanwhile the timing constraints of the  $m-ary$  LES  do not violate the timing constraints of the $n-ary$ ES, we call the $m-ary$ LES  is the subset of the $n-ary$  LES.
\end{LES}

For example, for a $3-ary$  LES $(A, B, C)$, its $2-ary$ subsets include $(A, B)$, $(A, C)$ and $(B, C)$. However, $(B, A)$ is not its $2-ary$ subset, since it violated the timing constraints of $(A, B, C)$.

\begin{LES}
\emph{$(n-1)-ary$ adjacent subset of $n-ary$  LES}: For an $(n-1)-ary$  LES that is the subset of an $n-ary$ LES, if all adjacent events of $(n-1)-ary$  LES are also adjacent in the $n-ary$  LES, we call the $(n-1)-ary$  LES the adjacent subset of the $n-ary$  LES.
\end{LES}

For an $n-ary$  LES $(a1, a2, \ldots, a(n-1), a(n))$, it has two $(n-1)-ary$ adjacent subsets: $(a1, a2, \ldots, a(n-1))$ and $(a2, \cdots, a(n-1), a(n))$.


\section{ SYSTEM DESIGN} \label{sec£ºsystem_design}



\subsection{ LogMaster architecture}


LogMaster includes three major components: \emph{Log agent}, \emph{Log server}, and \emph{Log database}.
   On each node, Log agent collects, preprocesses logs, and filters repeated events and periodic events. And then Log agent sends events to Log server for mining event rules, which are stored in Log database.

\begin{figure}[!htbp]
  \begin{center}
   \includegraphics[scale=0.5]{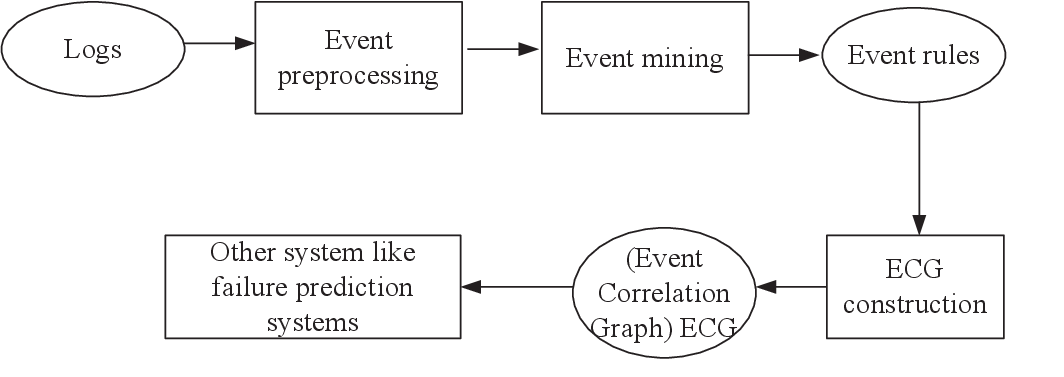}
   \end{center}
   \caption{\label{fig:approaches}The summary of event correlation mining approaches.}
   \end{figure}

   In Section ~\ref{subsec:correlation_mining_algorithms}, we will
  propose two event correlation mining approaches. At the same time, Log server constructs a set of graphs - event
     correlation graphs (ECG) to represent event correlations. Section ~\ref{subsec:ECGs}
     will introduce the details of ECG.
   LogMaster will mine event rules and their presentations
   - ECG for other systems, for example a failure prediction or fault diagnose system.
   Fig.~\ref{fig:approaches} summarizes our event correlation mining approaches.

\subsection {An metrics for measuring event correlations} \label{metrics}

In this subsection, we define a simple metrics to measure event correlations.

We assume that time  synchronization services are deployed on large-scale systems. The largest clock skew is
   easy to be estimated using a simple clock synchronization algorithm \cite{Gusella_TOSE89} \cite{Cristian_DC89}
   or the ntptrace \cite{ntptrace} tool. With a time synchronization service, like NTP, we can ignore the effect of
    clock skew in our algorithm, since a NTP service can guarantee time synchronization to a large extent.
    For example, NTPv4 can achieve an accuracy of 200 microseconds or better in local area networks under ideal conditions.



\begin{figure}[!htbp]
   \begin{center}
   \includegraphics[scale=0.5]{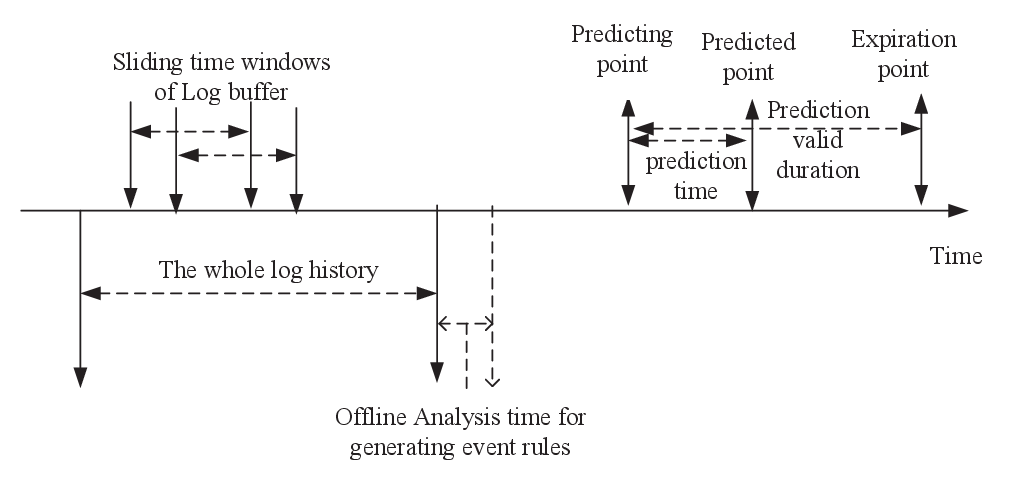}
   \end{center}
    \caption{The time relations in our event correlation mining and event prediction systems.}
    \label{fig:time-relation}
   \end{figure}

   As shown in Fig.~\ref{fig:time-relation}, we analyze the whole log history to generate event rules. In our approach, we use a sliding time window to analyze logs. For each current event, we save events within a sliding time window
(according to timestamps) to the log buffer, and analyze events in the
log buffer to mine event rules. After an event log has been
analyzed, we will advance the sliding time window according
to the timestamp of the current event.



Considering the events may happen interleavedly, we propose a \emph{confidence} metrics to measure the correlation of an LES as follows:

we count on two important attributes : \emph{the support count}, and \emph{the posterior count}.
The support count is the recurring times of the preceding events which are followed by the posterior event,
while the posterior count is the recurring times of the posterior event which follows the preceding events.
For example, if an event sequence $BACBBA$ occur in a time window, for a $2-ary$  LES $(A, B)$, the support count is one, and the posterior count is two; for a $3-ary$  LES $(A, C, B)$, the support count is one and the posterior count is two. The confidence metrics is calculated to measure the event correlation according to Equation \ref{confidence_equation}.


\begin{equation} \label{confidence_equation}
Confidence =  \frac{support \: count ( LES)}{posterior \: count ( LES)}
\end{equation}


According to Equation \ref{confidence_equation}, in $BACBBA$,  the confidence of an $2-ary$  LES $(A, B)$ is 1/2.
 In other words, if an event $A$ occurs, an event $B$ will occur with the probability of 50\%.


Based on the above definitions, we formally define \emph{frequent LES} and \emph{event rule} as follows: for an LES, if its adjacent subsets are frequent and its support count exceeds a predefined threshold, we call it  a frequent  LES.  For a frequent  LES, if its confidence exceeds a predefined threshold, we call it  an event rule.


\subsection{Event correlation mining algorithms} \label{subsec:correlation_mining_algorithms}

In this section, we propose two event correlation mining algorithms: an Apriori-LES algorithm and its improved version---Apriori-simiLES.

The notations of the Apriori-LES and Apriori-simiLES algorithm are listed in TABLE.~\ref{Algorithm_Notations}.
 \begin{table}[hbtp]
\renewcommand{\arraystretch}{1.3}
\centering
\begin{tabular}{|l|p{5.2cm}|}
  \hline
  \itshape Notation& \itshape Description \\ \hline
  \itshape Tw& \itshape the size of sliding time window  \\ \hline
  \itshape Sth& \itshape the threshold of the support counts of LES  \\ \hline
  \itshape Cth& \itshape the threshold of confidence of event rules \\ \hline
  \itshape C(k)& \itshape a set of frequent k-ary LES candidates  \\ \hline
  \itshape F(k)& \itshape a set of frequent k-ary LES \\ \hline
  \itshape R(k)& \itshape a set of k-ary event rules \\ \hline

\end{tabular}
\caption{Notations of the Apriori-LES and Apriori-simiLES}
\label{Algorithm_Notations}
\end{table}	


\subsubsection{Apriori-LES algorithm}

In data mining approaches, Apriori is a classic algorithm for learning association rules in transactional data \cite{Agrawal_VLDB94} \cite{Agrawal_SIGMODE93}.
Apriori uses a breadth-first search and a tree structure to count item set candidates. According to the downward closure lemma \cite{Agrawal_VLDB94} \cite{Agrawal_SIGMODE93}, a $k-length$ item set candidate contains all frequent $(k-1)-length$ item sets. The Apriori algorithm generates frequent $k-length$ item set candidates from frequent $(k-1)-length$ item sets.  After that, it scans transaction data to determine frequent item sets among the candidates.

As mentioned in Section \ref{introduction}, logs are significantly different from transaction data.
  We propose an algorithm to mine event rules, which we call the Apriori-LES algorithm. The Apriori-LES algorithm is as follows:

\emph{Step 1}: Predefine two threshold values £º$Sth$ and $Cth$ for the support count and the confidence, respectively.

\emph{Step 2}: Add all events that have different log IDs with the support count above the threshold value $Sth$ to $F(k=1)$;

\emph{Step 3}: $K=k+1$; $C(k) = \{\}$; $F(k) = \{\}$; $R(k) = \{\}$;

\emph{Step 4}: Get all frequent $k-ary$ LES candidates.
We generate the frequent $k-ary$ LES candidate by the \emph{LINK operation} of two frequent $(k-1)-ary$ adjacent subsets, and add it into $C(k)$.

The LINK operation is defined as below: for two frequent (k-1)-ary LES, if the last (k-2) log IDs of the one (k-1)-ary LES are same like the first (k-2) log IDs of the other (k-1)-ary LES, the result of the LINK operation is:
$LINK((a1, a2, \cdots, a(k-1)), (a2, \cdots, a(k-1), a(k)) = (a1, a2, \cdots, a(k-1), a(k))$

For example, if $(A, B,C)$ and $(B, C,D)$ are frequent $3-ary$ LES in $F(3)$, then a $4-ary$ LES $(A, B, C, D)$ is a
frequent $4-ary$ LES candidate, which we will add into $C(4)$.

\emph{Step 5}: Scan the logs to validate each frequent $k-ary$ LES candidates in $C(k)$. The support count and posterior count of each $k-ary$ LES candidate in $C(k)$ is counted. For a frequent $k-ary$ LES candidate
$(a1,a2,\cdots,a(k-1),a(k))$, if event $a(k-1)$ occurs after any event in $(a1,a2,\cdots,a(k-2))$ and before the posterior event $m$ times, and the posterior event occurs after any event in $(a1,a2,\cdots,a(k-2), a(k-1))$ $n$ times, we increment the support count and the posterior count of the $k-ary$ LES candidate by $m$ and $n$, respectively.

\emph{Step 6}: Generate frequent $k-ary$ LES and $k-ary$ event rules. For a $k-ary$ frequent LES candidate, if its support count is above the threshold $Sth$, add it into $F(k)$. For a $k-ary$ LES candidate, if
its support count and confidence are above the threshold values: $Sth$ and $Cth$, respectively,  add it into $R(k)$.


\emph{Step 7}: Loop until all frequent LES and event rules are found, and save them in Log database. If $R(k)$ is not null, save
$k-ary$ event rules in $R(k)$. If $F(k)$ is not null, go to \emph{step 3}; else end the algorithm.

\subsubsection{Apriori-simiLES algorithm}

As shown in Section \ref{sec:evaluation}, the Apriori-LES algorithm still suffers from inefficiency, and generates a large  amount of frequent LES candidate, which may lead to a long analysis time. In order to improve performance and save costs while  ensuring the algorithm's efficiency, we observe the breakdown of event rules through mining about ten days's logs of the Hadoop system (in Nov 2008) and the HPC cluster system (in Jan 2004 ), respectively.

We set the following configuration in the Apriori-LES algorithm: the sliding time window ($Tw$), the support count threshold ($Sth$), and the confidence threshold ($Cth$) are 60 minutes, 5, and 0.25, respectively. We only mine $2-ary$ event rules so as to simplify the experiments.

In all, we get 517 $2-ary$ event rules in the Hadoop logs and 156 $2-ary$ event rules in the HPC logs. When we analyze the breakdown of 2-ary event rules generated by the Apriori-LES algorithm, we find that most of $2-ary$ event rules are composed of events that occur on the same nodes or the same
applications, or have the same event types. This phenomenon is probably due to: (a) error may spread in a single node. For example: one application or process error can lead to another application or process error. (b) replicated applications in multiple nodes may have same errors or software bugs, and same failure events may appear in multiple nodes. (c) nodes in an large-scale system need to transfer data and communicate with each other, so a failure on one node may cause failures of same event types on other nodes. (d) a failure on one node may change the cluster system environment, which may cause failures of same event types on other nodes.
The analysis results are shown in TABLE.~\ref{Event Rules Spread_Hadoop} and
TABLE.~\ref{Event Rules Spread_HPC}.

  \begin{table}[hbtp]
\renewcommand{\arraystretch}{1.3}
\centering
\begin{tabular}{|l|p{0.7cm}|p{1.0cm}|p{1.7cm}|p{1.7cm}|}
  \hline
   \itshape Description & \itshape All  &\itshape Same nodes &\itshape Same event types &\itshape Same applications \\ \hline
   \itshape count& \itshape 517  &\itshape 168 &\itshape 172 &\itshape 159 \\ \hline
   \itshape Percent(\%)& \itshape 100\% &\itshape 32.5\% &\itshape 33.3\% &\itshape 30.8\% \\ \hline
\end{tabular}
\caption{The breakdown of $2-ary$ event rules in the Hadoop logs.}
\label{Event Rules Spread_Hadoop}
\end{table}

  \begin{table}[hbtp]
\renewcommand{\arraystretch}{1.3}
\centering
\begin{tabular}{|l|p{0.7cm}|p{1.0cm}|p{1.7cm}|p{1.7cm}|}
  \hline
   \itshape Description & \itshape All  &\itshape Same nodes &\itshape Same event types &\itshape Same applications \\ \hline
  \itshape count& \itshape 156  &\itshape 10 &\itshape 48 &\itshape 52 \\ \hline
  \itshape Percent(\%)& \itshape 100\% &\itshape 6.4\% &\itshape 30.8\% &\itshape 33.3\% \\ \hline
\end{tabular}
\caption{The breakdown of $2-ary$ event rules in the HPC logs.}
\label{Event Rules Spread_HPC}
\end{table}

							
On the basis of these observations, we propose an improved version of the Apriori-LES algorithm: Apriori-simiLES.
The distinguished difference of Apriori-simiLES from Apriori-LES is that the former uses an event filtering policy before
the event correlation mining. The event filtering policy is described as below: to reduce the number of the analyzed events and decrease the analysis time, we only analyze correlations of events that occur in (a) \emph{the same nodes} or (b) \emph{the same applications}, or have (c) \emph{the same event types}.

The Apriori-simiLES algorithm includes two rounds of analysis: a single-node analysis and a multiple-node analysis. In the single-node analysis, we use the Apriori-LES algorithm to mine event rules that have same $node \: IDs$. And in the multiple-node analysis,
we use the Apriori-LES algorithm to mine event rules that are of the same application names or the same event types but with different $node \: IDs$.

\subsection{ECGs construction}  \label{subsec:ECGs}

After two rounds of analysis, we get a series of event rules. Based on the event rules, we propose a new abstraction---event correlation graphs (ECGs) to represent event rules.

A ECG is a directed acyclic graph (DAG). A vertex in a ECG represents a event. 
For a vertex, its children
  vertexes are its posterior events in event rules. There are two types of vertexes: dominant and recessive. For a $2-ary$ event
  rule, such as $(A, B)$, the vertexes representing events $A$, $B$ are dominant vertexes. 
  An additional vertex $A \land B$ represents the case that $A$ and $B$ occurred and
  $B$ occurred after $A$. The vertex $A\wedge B$ is a recessive vertex.

Vertexes are linked by edges. An edge represents the correlation of two events linked by the edge.
Each edge in ECG has five attributes: \emph{head vertex}, \emph{tail vertex}, \emph{support count}, \emph{posterior count} and \emph{edge type}.
Similar to vertexes, edges have two types: \emph{dominant} and \emph{recessive}.  If two vertexes linked by an edge are dominant,
the edge type is dominant; otherwise, the edge type is recessive. For a $2-ary$ event rule $(A, B)$, the head vertex is $B$,
and the tail vertex is $A$; the support count and posterior count of the edge is the support count and posterior count
of the rule event $(A, B)$, respectively. For a $3-ary$ event rule $(A, B, C)$, the head vertex is $C$, and the tail vertex is the recessive
vertex $A \wedge B$; the support count and the posterior count of the edge is the support count and
the posterior count of the event rule $(A, B, C)$, respectively.

\begin{figure}[!htbp]
   \begin{center}
   \includegraphics[scale=0.8]{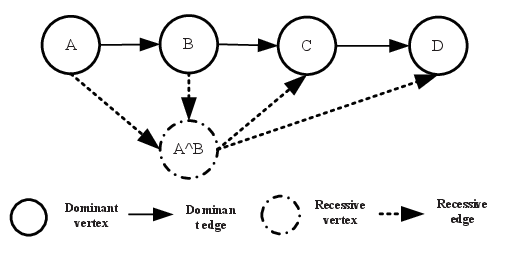}
   \end{center}
   \caption{An example of ECG.}
   \label{fig:ECG-example}
   \end{figure}

An example is shown in Fig.~\ref{fig:ECG-example}. There are three $2-ary$ event rules: $(A, B)$, $(B, C)$ and $(C, D)$, and
two $3-ary$ $(A, B, C)$ and $(A, B, D)$. We generate four dominant vertexes, which represent events $A$, $B$, $C$, and $D$. We also
generate three dominant edges which represent the event rules $(A, B)$ and $(B, C)$ and $(C, D)$. In additional, we also generate
a recessive vertex ($A\wedge B$). A recessive edge ($A\wedge B \rightarrow C$) represents the event rule $(A, B, C)$, and a
recessive edge ($A\wedge B \rightarrow D$) represents the event rule $(A, B, D)$. The additional vertex $A\wedge B$ is the
child of both $A$ and $B$.

Choosing the ECG abstraction has three reasons: first, the visualized graphs are easy to understand by system managers and operators; second, the abstraction facilitates modeling sophisticated correlations of events, such as $k-ary$ ($k>2$) event rules; third, ECGs can be easily updated in time since the attributes of edges and vertexes are easily updated when a new event comes.

The construction of ECGs includes three main steps:

\emph{Step 1}: Construct a group of ECGs based on event rules found in the single-node analysis. Each ECG represent
correlations of events in one node.

Based on event rules generated in the analysis, the vertexes and edges of ECGs are created. For each event rule,
dominant vertexes and recessive vertexes are generated, and edges between vertexes are created too.

\emph{Step 2}: ECGs that represent correlations of events on multiple nodes are constructed based on event rules found in multiple-nodes analysis.

\emph{Step 3}: The index of ECGs is created, and the positions of events in ECGs are also saved. We can locate events by using these indexes.

The ECG ID and the ECG entrance vertex are the index of the ECGs. The ECG position is the index of each event in an ECG. So it is convenient to locate events in the ECGs.

After these three steps, a series of ECGs that describe the correlation of events are constructed.

\subsection{Event Prediction }  \label{subsec:prediction}
For each prediction, there are three important timing points: \emph{predicting point},
 \emph{predicted point}, and \emph{expiration point}. The relations of those timing points are shown in Fig.~\ref{fig:time-relation}.

  The prediction system begins predicting events at the timing
of the predicting point. The predicted point is the occurrence timing of the predicted event. The expiration point refers to the expiration
 time of a prediction, which means this prediction is not valid if the actual occurrence timing of the event passed the expiration point.

 In addition, there are two important derived properties for each prediction: \emph{prediction time}, and \emph{prediction valid duration}.
The prediction time is the time difference between the predicting point and the predicted point, which is the time span left for system administrators to respond with the possible upcoming failures. The prediction valid duration is the time difference between the predicting point and the expiration point.

The event prediction algorithm based on ECGs is as follows:

\emph{Step 1}: Define the prediction probability threshold $Pth$, and the prediction valid duration $Tp$.

\emph{Step 2}: When an event comes, the indexes of events are searched to find matching ECGs and
the corresponding vertexes in ECGs. The searched vertexes are marked. For a recessive edge, if its tail vertex is marked,
the recessive edge is marked, too. For a recessive head vertex, if all recessive edges are marked, the recessive vertex is
 also marked. We mark vertexes so as to predict events; and the head vertexes that are dominant vertexes are searched
 according to the edges (both dominant and recessive edge types) linked with marked vertexes.

 \emph{Step 3}: The probabilities of the head vertexes are calculated according to the attributions of vertexes that
 are marked and their adjacent edges in the ECGs. For a head vertex, we calculate its probability as the probability of
 tail vertex times the confidence of the edge. If a head vertex is linked with two marked vertexes with different edges,
 we will calculate two probabilities, and we choose the largest one as the probability of the head vertex.

\begin{figure}[htbp]
   \begin{center}
   \includegraphics[scale=0.7]{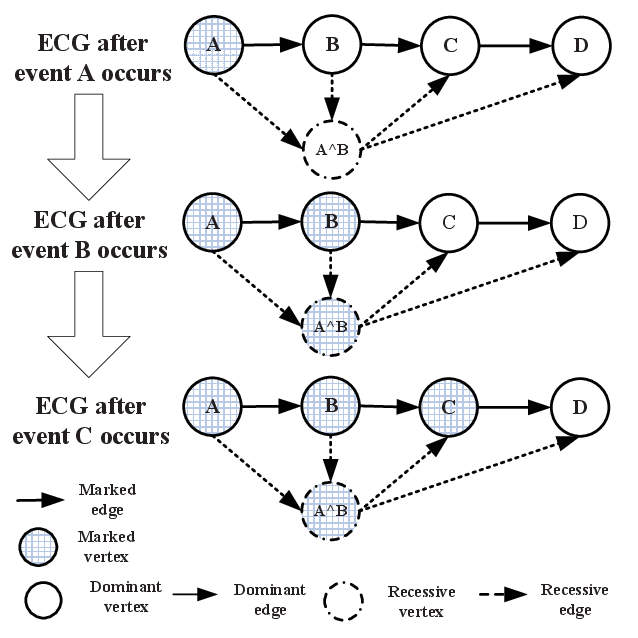}
   \end{center}
   \caption{An example of event prediction.}
   \label{fig:ECG-prediction}
   \end{figure}

An example is shown in Fig.~\ref{fig:ECG-prediction}. When an event $A$ occurs, the vertex $A$ and the edge $A\rightarrow A\wedge B$
are marked. We can calculate the probability of $B$ according to the confidence of the edge $A\rightarrow B$. The probability of
event $C$ also can be calculated by the dominant edges $A\rightarrow B$ and $B\rightarrow C$. If the probability of event $B$ or
 event $C$ is above the prediction probability threshold $Pth$, it is predicted. So does the event $D$.

\begin{eqnarray}
Probability(B)=confidence(A \rightarrow B)
\end{eqnarray}
\begin{eqnarray}
Probability(C)=probability(B) \ast\: confidence(B \rightarrow C)
                              \nonumber\\
=confidence(A \rightarrow B) \ast confidence(B \rightarrow C)
\end{eqnarray}

When an event $B$ occurs later, the vertex $B$ and the edge $B\rightarrow A\wedge B$ are marked. Because the edge $A\rightarrow A\wedge B$
and $B\rightarrow A\wedge B$ are both marked, so the vertex $A\wedge B$ are marked too. The probability of events $C$ and $D$ are also calculated according to the new edges.

\emph{Step 4}: If the probability of a head vertex is above the prediction probability threshold $Pth$, then the head vertex is
the predicted event.

\emph{Step 5}: Loop \emph{Step 2}.

\section{LogMaster Implementation}  \label{sec:implementation}

We have implemented an event correlation mining system and an event prediction system. The main modules of LogMaster are as follows:

(1) The configuration file and Log database. A XML-format configuration file named config.xml includes the regular expressions of important definitions, formats, and keywords that are used to parse logs and the configuration of Log database.

Log database is a MySQL database, including the tables of the formatted logs, the filtered logs, and the event rules. We define event attributions in the section "definitions", and the definitions include $timestamp$, $node \: name$, $application$, $process$, $process \: id$, $user$, $description$, and so on. The formats of event logs are defined in the section "format". All keywords that are used to decide the severity degree and event type of logs are defined in the section "keywords".

 (2) The Python  scripts are used to parse system log files according to the regular expressions defined in the configuration file. The parsed logs are saved to the table of the formatted logs in Log database.

An example of the original event logs and the corresponding formatted event logs is shown as bellows:

[\textbf{Original event log}] Oct 26 04:04:20 compute-3-9.local smartd [3019]: Device: /dev/sdd, FAILED SMART self-check. BACK UP DATA NOW!

[\textbf{Formatted event log}] timestamp ="20081024040420", node name ="compute-3-9.local", format ="format1", keyword = "FAILED SMART", application="smartd", process="NULL", process id="3019", user="NULL", description="Device: /dev/sdd, FAILED SMART self-check. BACK UP DATA NOW!".

(3) Log server is written in Java. The simple filtering operations are performed to remove repeated events and periodic events.
The filtered logs are saved in the table of filtered logs in Log database. We implemented both Apriori-LES and Apriori-semiLES algorithms. The attributions of each event rules, including event attributions, support count, posterior count, and confidence are saved to the table of event rules in Log database.


(4) We implement the event rules based prediction system  in C++.

\section{Evaluations} \label{sec:evaluation}

In this section, we use LogMaster to analyze logs of a production Hadoop cluster system and a HPC cluster system logs,
the detail of two logs are described in Section ~\ref{sec:background}.
 The server we used has two Intel Xeon Quad-Core E5405 2.0GHZ processors, 137GB disk, and 8G memory.

\subsection{ Filtering Events}

 In this step, we remove repeated events and periodic events.


 After collecting logs, Log sever will perform a simple filtering according to two observations: first, there are two types of repeated events:  one kind of repeated events are recorded by different subsystems, and the other kind of repeated events  repeatedly occur in a short time windows. Second, periodic
events. Some events periodically occur with a fixed interval
because of hardware or software bugs. Each type of periodic
events may have two or more \emph{fixed cycles}. For example, if a daemon in a node monitors CPU or memory systems periodically,
it may produce large amount of periodic events.

We use a simple statistical algorithm and a simple clustering algorithm to remove repeated events and
periodic events, respectively. The solution to removing repeated events is as
follows: for the first step, we treat events with the same $log \:   ID$ and the same $timestamp$
as repeated events. For the second step, we treat events with the same $log \:   ID$ occurring in a small time windows as repeated events.
In this experiment, we set this interval threshold as 10 seconds, and the reason is that we consider the repeated events should occur in a short time windows.

The solution to removing periodic events is as follows: for each log ID, update the counts of events
for different intervals. For periodic events with the same log ID, if
the count and the event percent of the same interval, which is obtained against all periodic
events with the same log ID,  is
higher than the predefined threshold values, respectively, we
consider the interval as a fixed cycle. In our experiment, we
set two predefined threshold values of the Hadoop logs and the HPC logs as (20, 0.2) and (20, 0.1),  respectively. The effects of different threshold values on the number of filtered events can be found at  Appendix \ref{appendix_preprocessing}.
Lastly, we only keep one event for periodic events with the same fixed cycle. For
periodic events, only events deviated from the fixed cycle are reserved.


 TABLE~\ref{filtering-result} shows the experiments results. In preprocessing, our python scripts parse about 977,858 original Hadoop event entries in 4 minutes 24 seconds, and interpret those events into nine-tuples, which are stored into the MySQL database. We parse 176,043 original HPC cluster event entries in 2 minutes 28 seconds. Please note that we only select the node logs from the HPC logs without including other events, e.g., that of "switch module" , since the Hadoop logs only include node logs.

 \begin{table}[hbtp]
\renewcommand{\arraystretch}{1.3}
\centering
\begin{tabular}{|l|p{0.7cm}|p{1.0cm}|p{1.05cm}|p{1.05cm}|p{1.2cm}|}
  \hline
  \itshape logs& \itshape raw logs &\itshape Pre processing &\itshape Removing repeated events &\itshape Removing periodic events &\itshape Compression rate \\ \hline
  \itshape Hadoop& \itshape 977,858 &\itshape 977,858 &\itshape 375,369 &\itshape 26,538&\itshape 97.29\%	 \\ \hline
  \itshape HPC cluster&\itshape 433,490 &\itshape 176,043  &\itshape 152,112 &\itshape 132,650 &\itshape 69.4\% \\ \hline
\end{tabular}
\caption{The results of preprocessing and filtering logs.}
\label{filtering-result}
\end{table}

In this experiment, the compression rate of the Hadoop logs can achieves 97.29\%,  but the compression rate of the HPC logs only achieves 69.4\%. The reason is probably that the Hadoop logs have a large amount of repeated events, while  the HPC logs have relatively small number of events for a long period, and hence have less repeated events. The periodic events of HPC logs are small, and the reason is that the HPC logs have a long time span.

\subsection{ Comparison of two event correlation mining algorithms} \label{comparision}

In this step, we compare two proposed algorithms: Apriori-LES and Apriori-simiES.    We use (a) \emph{the average analysis time per events} and (b) \emph{the number of event rules} to
   evaluate the computational complexity and the efficiency of Apriori-LES and Apriori-simiES algorithms, respectively.

For the Hadoop logs, we analyze 43 days' logs from  2008-10-26 04:04:20.0 to 2008-12-09 23:21:28.0.  For the HPC logs,
we analyze 48 days' logs from  2003-12-26 22:12:30.0  to 2004-02-13 03:02:39.0.

Before reporting experiment results of two algorithms, we pick the following parameters as the
baseline configuration of LogMaster for comparisons. Through
comparisons with an large amount of experiments, we set the
baseline parameters in LogMaster: Hadoop logs---[Tw60/Sth5/Cth0.25] and HPC logs---[Tw60/Sth5/Cth0.25].
[Tw\emph{x}/Sth\emph{y}/Cth\emph{z}] indicates that the sliding time window $Tw$ is
$x$ minutes, the threshold of support count $Sth$ is $y$, and the
threshold of confidence $Cth$ is $z$. The effect of varying parameters ($Tw$, $Sth$ and $Cth$) on the average analysis time per event and the number of event rules can be found in Appendix \ref{appendix_parameters}.

  The comparison experiments show that: for the Hadoop logs, the average analysis time of Apriori-simiLES is about 10\%-20\% of that of Apriori-LES, while Apriori-simiLES obtains about 60\%-70\% event rules of that of Apriori-LES; For the HPC logs, the average analysis time of Apriori-simiLES is about 10\%-20\% of that of Apriori-LES algorithm, while
Apriori-simiLES obtains about 80\%-90\% event rules of that of Apriori-LES.


\subsection{The summaries of events and event rules in two typical cluster systems } \label{rule_summary}

 In two typical cluster systems for Cloud and HPC, respectively, we give the summaries of the events and events rules, which are generated by the Apriori-LES algorithm with the baseline parameters mentioned above.

\begin{figure}[htbp]
   \begin{center}
   \begin{minipage}[c]{0.25\textwidth}
   \centering \includegraphics[width=4.0cm,height=3.0cm]{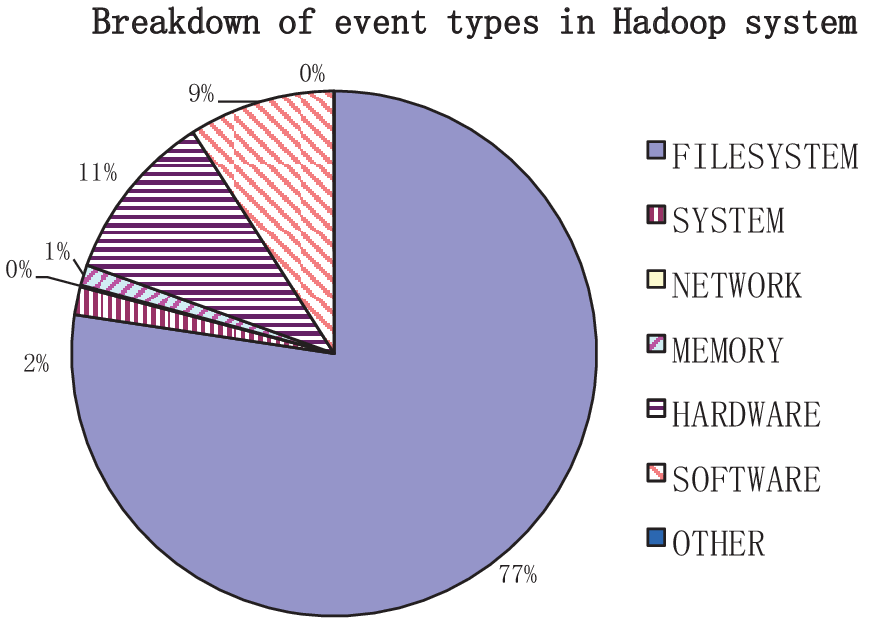}
   \end{minipage}%
   \begin{minipage}[c]{0.25\textwidth}
   \centering \includegraphics[width=4.0cm,height=3.0cm]{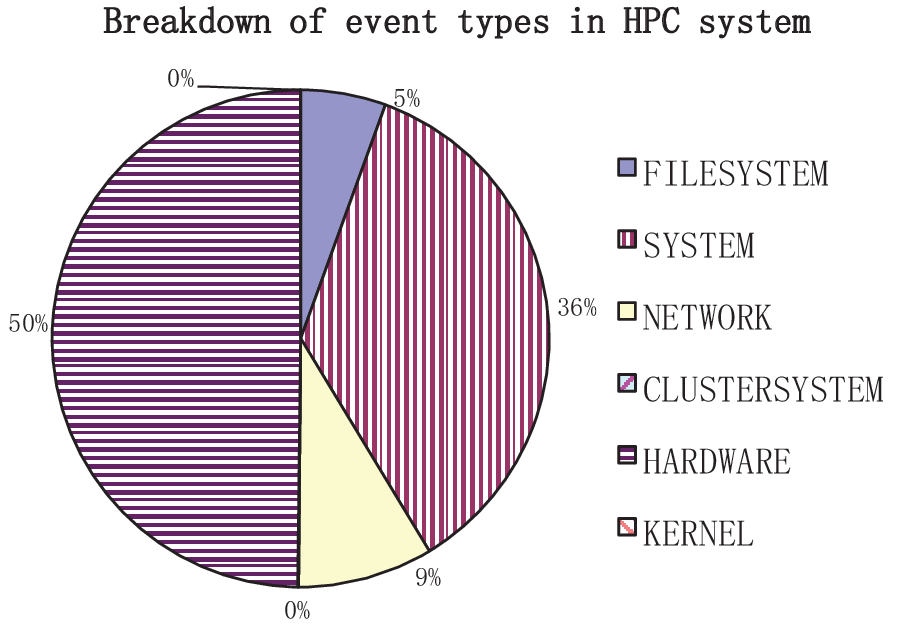}
   \end{minipage}
   \caption{\label{fig:event_spread}The breakdown of event types of two system logs.}
   \end{center}
   \end{figure}

(a) In the Hadoop logs, the number of events of different types ranks according to the order: FILESYSTEM, HARDWARE, SOFTWARE, SYSTEM, MEMORY, NETWORK and OTHER. In the HPC logs, the number of events of different types ranks according to the order: HARDWARE, SYSTEM, NETWORK, FILESYSTEM, CLUSTERSYSTEM, and KERNEL. Please note that the event types of two logs are slightly different. For the HPC logs, the event types are recorded in the original logs, while for the Hadoop logs the event types are parsed by ourself. The breakdown of logs is shown in Fig.~\ref{fig:event_spread}.

(b) Most of event rules are composed of events with the FILESYSTEM type in the Hadoop logs, and the reason
may be that applications running on the Hadoop cluster system are data-intensive, and need to access the
file system frequently; However, most of event rules in the HPC logs are composed of events of HARDWARE  and SYSTEM types, and it is probable that hardware and system level errors more easily lead to failures in HPC cluster systems.

With the same baseline configuration like that in \ref{comparision}, we obtain 2 or 3-ary event rules with Apriori-LES and Apriori-simiLES as shown in TABLE~\ref{event-rules}.

\begin{table}[hbp]
\begin{tabular}{|l|p{1.2cm}|p{1.2cm}|p{1.2cm}|p{1.2cm}|}
\hline
\multirow{2}{*}{Event Rules} & \multicolumn{2}{|c|}{Hadoop logs} & \multicolumn{2}{|c|}{HPC cluster logs} \\
\cline{2-5}
\itshape  &\itshape Apriori-LES &\itshape Apriori-simiLES & \itshape Apriori-LES &\itshape Apriori-simiLES \\ \hline
\itshape $2-ary$ &\itshape 2413 &\itshape 1520 & \itshape 4726 &\itshape 3990 \\ \hline
\itshape $3-ary$ &\itshape 1603 &\itshape 1603 & \itshape 1695 &\itshape 1695 \\ \hline
\end{tabular}
\caption{The event rules obtained from two algorithms.}
\label{event-rules}
\end{table}

We also analyze the breakdown of event rules that lead to failure events. As shown  in TABLE~\ref{failure-rules}, with the exception of the event rules in the HPC logs obtained with the Apriori-simiLES algorithm, event rules that identify correlations between non-failure events (including "INFO", "WARNING", "ERROR")  and failure events ("FATAL" and "FAILURE")) dominate over the event rules that identify the correlations between different failure events.

\begin{table}[hbp]
\begin{tabular}{|l|p{1.0cm}|p{1.1cm}|p{1.0cm}|p{1.1cm}|}
\hline
\multirow{2}{*}{Failure Rules(No.)} & \multicolumn{2}{|c|}{Hadoop logs} & \multicolumn{2}{|c|}{HPC cluster logs} \\
\cline{2-5}
\itshape  &\itshape Apriori-LES &\itshape Apriori-simiLES & \itshape Apriori-LES &\itshape Apriori-simiLES \\ \hline
\itshape Configuration &\multicolumn{2}{|c|}{[Tw60/Sth5/Cth0.25]} &\multicolumn{2}{|c|}{[Tw60/Sth5/Cth0.25]} \\   \hline
\itshape Non-failures$\rightarrow$Failures&\itshape 377 &\itshape 180 & \itshape 97 &\itshape 5 \\ \hline
\itshape Failures$\rightarrow$Failures &\itshape 86 &\itshape 78 & \itshape 26  &\itshape 26 \\ \hline
\end{tabular}
\caption{The event rules obtained from two algorithms}
\label{failure-rules}
\end{table}

 \begin{table}[htbp]
\begin{tabular}{|l|p{0.7cm}|p{0.9cm}|p{0.9cm}|p{0.9cm}|p{0.9cm}|p{1.0cm}|}
\hline
\itshape logid1&\itshape logid2&\itshape nodeid1 &\itshape type1 &\itshape nodeid2 &\itshape type2 &\itshape confidence  \\ \hline
\itshape 3314 &\itshape 3311 &\itshape 249 & \itshape memory &\itshape 249 & \itshape system &\itshape 0.997487 \\ \hline
\itshape 370 &\itshape 359 &\itshape 42 & \itshape hardware  &\itshape 42 & \itshape filesystem &\itshape 0.993789 \\ \hline
\itshape 91 &\itshape 89 &\itshape 4 & \itshape software &\itshape 4  & \itshape system &\itshape 0.961538 \\ \hline
\itshape 2034 &\itshape 2035 &\itshape 164 & \itshape filesystem &\itshape 164 & \itshape filesystem &\itshape 0.952381  \\ \hline
\itshape 1412 &\itshape 1413 &\itshape 120 & \itshape software &\itshape 120 & \itshape software &\itshape 0.947368  \\ \hline
\itshape 66 &\itshape 64 &\itshape 2 & \itshape system &\itshape 2 & \itshape software &\itshape 0.947368  \\ \hline
\itshape 147 &\itshape 148 &\itshape 12 & \itshape filesystem &\itshape 12 & \itshape hardware &\itshape  0.9375   \\ \hline
\itshape 3632 &\itshape 3628 &\itshape 260 & \itshape system &\itshape 260 & \itshape software &\itshape  0.933333  \\ \hline
\itshape 3627 &\itshape 3628 &\itshape 270& \itshape  filesystem &\itshape 270 & \itshape software  &\itshape  0.933333  \\ \hline
\itshape 172 &\itshape 169 &\itshape 22 & \itshape hardware &\itshape 22 & \itshape filesystem &\itshape  0.928571   \\ \hline
\end{tabular}
\caption{Top 10 $2-ary$ event rules in the order of the descending confidence in the Hadoop logs}
\label{Hadoop_top10}
\end{table}

  \begin{table}[htbp]
\begin{tabular}{|l|p{0.7cm}|p{0.9cm}|p{0.9cm}|p{0.9cm}|p{0.9cm}|p{1.0cm}|}
\hline
\itshape logid1&\itshape logid2&\itshape nodeid1 &\itshape type1 &\itshape nodeid2 &\itshape type2 &\itshape confidence  \\ \hline
\itshape 4671 &\itshape 4682 &\itshape 260 & \itshape hardware &\itshape 260& \itshape hardware &\itshape 0.975 \\ \hline
\itshape 2598 &\itshape 2580 &\itshape 153  & \itshape hardware &\itshape 153 & \itshape hardware &\itshape 0.975 \\ \hline
\itshape 2601 &\itshape 2619 &\itshape 154 & \itshape hardware &\itshape 154 & \itshape hardware &\itshape 0.928571 \\ \hline
\itshape 2193 &\itshape 2180 &\itshape 131 & \itshape hardware  &\itshape 131 & \itshape hardware &\itshape 0.923077  \\ \hline
\itshape 2774 &\itshape 2883 &\itshape 162 & \itshape hardware &\itshape 167 & \itshape hardware &\itshape 0.923077  \\ \hline
\itshape 2796 &\itshape 2883 &\itshape 163 & \itshape  hardware &\itshape 167 & \itshape hardware &\itshape  0.923077  \\ \hline
\itshape 2819 &\itshape 2985 &\itshape 164 & \itshape hardware &\itshape 172 & \itshape hardware &\itshape  0.923077   \\ \hline
\itshape 2556 &\itshape 2539 &\itshape 151 & \itshape hardware  &\itshape 151 & \itshape hardware &\itshape 0.916667 \\ \hline
\itshape 3195 &\itshape 3212 &\itshape 182 & \itshape hardware &\itshape 182 & \itshape hardware &\itshape 0.916667  \\ \hline
\itshape 2661 &\itshape 2643 &\itshape 156 & \itshape hardware &\itshape 156 & \itshape hardware &\itshape 0.909091  \\ \hline
\end{tabular}
\caption{Top 10 $2-ary$ event rules in order of confidence in the HPC logs.}
\label{HPC_top10}
\end{table}

 The top 10 $2-ary$ event rules in the order of the descending confidence in the Hadoop logs and the HPC logs are shown in TABLE ~\ref{Hadoop_top10} and TABLE ~\ref{HPC_top10}, respectively.
For the top one $2-ary$ event rule in the Hadoop logs---$(3314, 3311)$, the original logs are shown in TABLE \ref{original_log}.

\begin{table}[hbtp]
\renewcommand{\arraystretch}{1.3}
\centering
\begin{tabular}{|p{8.0cm}|}
\hline
[Log id=3314] 2008-12-06 05:04:27 compute-12-9.local looks like a 64bit wrap, but prev!=new

[Log id=3311] 2008-12-06 05:04:57 compute-12-9.local c64 32 bit check failed  \\
\hline
\end{tabular}
\caption{The original logs of $(3314, 3311)$.}
\label{original_log}
\end{table}

\subsection {Evaluation of predication}

 After mining the event rules, we need to consider whether these event rules are suitable for predicting events. We evaluate our algorithms in three scenarios: (a) predicting all events on the basis of both failure and non-failure events; (b) predicting only failure events on the basis of both failure and non-failure events; (c)  predicting failure events after removing non-failure events.

 On the basis of event rules obtained in Section \ref{rule_summary}, we predict  21 days' logs from 2008-12-10 00:00:38.0 to 2008-12-31 15:32:03.0	in the Hadoop logs, and 14 days' logs from 2004-02-13 03:02:41.0 to 2004-02-27 19:02:00.0, respectively.

  We use \emph{the precision rate}, \emph{the recall rate}, and \emph{the average prediction time of event prediction} to evaluate the prediction.  \emph{The true positive (TP)} is the count of events which are correctly predicted. \emph{The false positive (FP)} is the count of events which are predicted but not appeared in the prediction valid duration. The precision rate is  the ratio of the correctly predicted events (TP) to all predicted events, including TP and FP. The recall rate is  the ratio of correctly predicted events (TP) to all filtered events. We calculate the average prediction time according to Equation \ref{eqa_average_prediction_time}.



\begin{eqnarray} \label{eqa_average_prediction_time}
 \textrm{The average prediction time}=   \nonumber\\
 \frac{\sum(\textrm{the predicted point}-\textrm{the predicting point})}{\textrm{count of all predicted events}}
\end{eqnarray}	



   There are two parameters that affect the prediction accuracy, including the prediction probability threshold ($Pth$) and the prediction     valid duration ($Tp$).     Before reporting experiment results, we pick the following parameters as the
baseline configuration for comparisons. Through comparisons
with large amount of experiments, we set the baseline parameters of the Hadoop logs-[Tw60/Sth5/Cth0.25/Pth0.5/Tp60] and the baseline parameters of the HPC logs-[Tw60/Sth5/Cth0.25/Pth0.5/Tp60]. Please note that $Tw$, $Sth$, $Cth$ just keep the same baseline parameters in Section \ref{comparision}.

[Pth\emph{u}/Tp\emph{v}] indicates that the prediction probability
threshold $Pth$ is $u$, and the prediction valid duration $Tp$ is $v$
minutes. The effects of varying $Pth$ and $Tp$ on failure predictions can be found at Appendix \ref{appendix_prediction_parameters}.

\begin{figure}[htbp]
   \begin{center}
   \begin{minipage}[c]{0.25\textwidth}
   \centering \includegraphics[width=4.6cm,height=3.5cm]{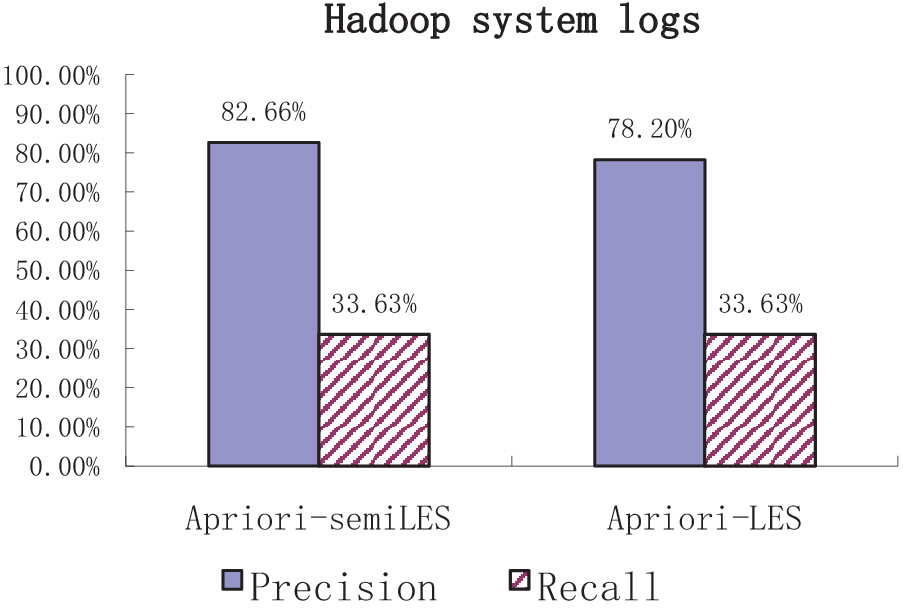}
   \end{minipage}%
   \begin{minipage}[c]{0.25\textwidth}
   \centering \includegraphics[width=4.6cm,height=3.5cm]{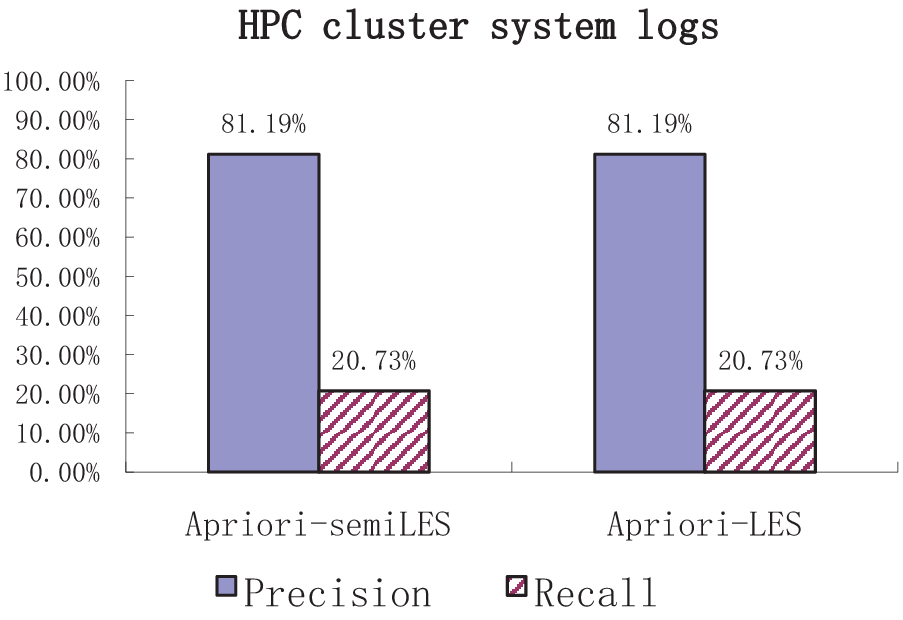}
   \end{minipage}
   \caption{The precision rate and recall rate of predicting events (including failure and non-failure events) on the basis of  failure and non-failure events. }
   \label{fig:event_prediction}
   \end{center}
   \end{figure}

First, on the basis of events of both failure and non-failure events, we predict all events, including "INFO", "WARNING", "ERROR", "FAILURE", and "FATAL" events.    The precision rates and recall rates of predicting events in the Hadoop logs and the HPC  logs are shown in Fig.~\ref{fig:event_prediction}. From Fig.~\ref{fig:event_prediction},
we can observe that with Apriori-LES, the precision rates of the Hadoop logs and the HPC logs are high as 78.20\%, 81.19\%, respectively, while the recall rates of the Hadoop logs and the HPC logs are 33.63\% and 20.73\%, respectively. The reason for the low recall rates is that we still keep rich log information after filtering events, including 26,538  entries (2.71\% of the original Hadoop logs) and 132,650 entries (30.6\% of the original HPC logs), respectively. We also notice that adopting a more efficient algorithm---Apriori-semiLES, which mines fewer event rules, results in higher precision rates. This is because with Apriori-LES we can obtain more event rules, which predicts more events which not happen.

\begin{figure}[htbp]
   \begin{center}
   \begin{minipage}[c]{0.25\textwidth}
   \centering \includegraphics[width=4.6cm,height=3.5cm]{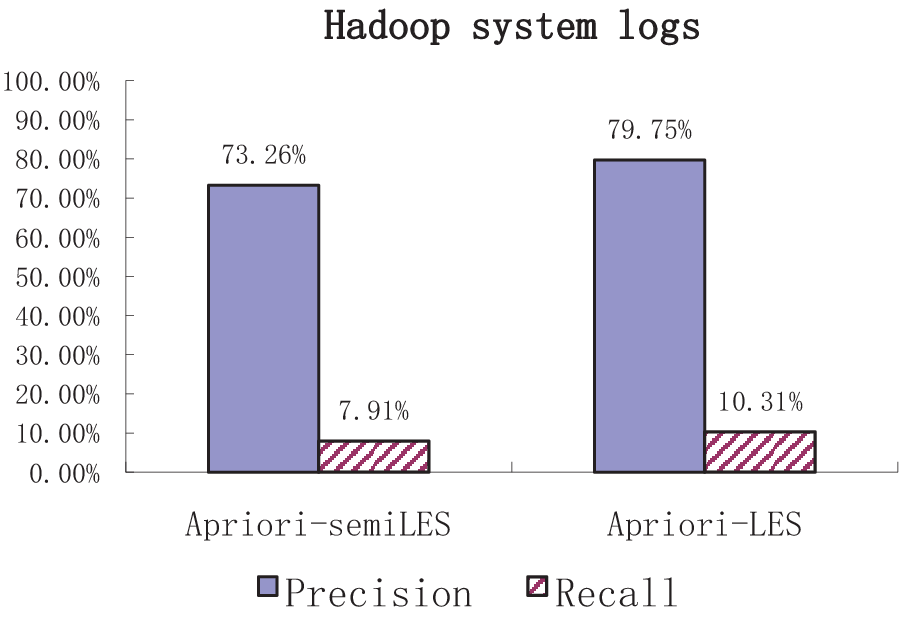}
   \end{minipage}%
   \begin{minipage}[c]{0.25\textwidth}
   \centering \includegraphics[width=4.6cm,height=3.5cm]{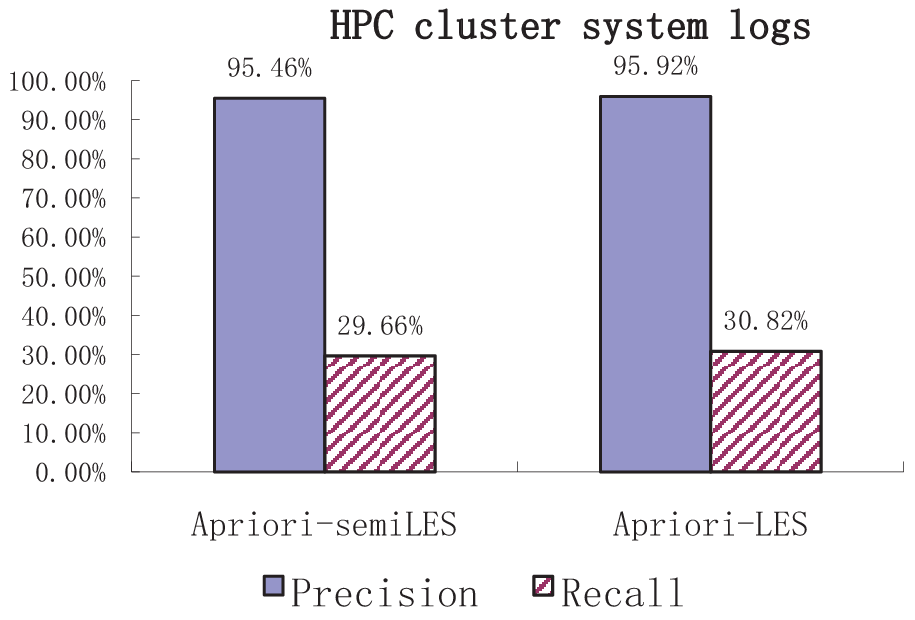}
   \end{minipage}
   \caption{The precision rate and recall rate of predicting \emph{failure} events on the basis of both failure and non-failure events.}
   \label{fig:failure_prediction}
   \end{center}
   \end{figure}

Second, on the basis of events of all types, we only predict failure events (Failure and FATAL types), of which the precision rates and the recall rates of two logs are shown in Fig.~\ref{fig:failure_prediction}.

For predicting failure events, Liang {\em et al.} \cite{Liang_DSN05} suggest removing events with the types of \emph{INFO, WARNING, and ERROR} in preprocessing events. Lastly, following their idea, we remove non-failure events, and then predict failure events in the subsequent experiments.

\begin{figure}[htbp]
   \begin{center}
   \begin{minipage}[c]{0.25\textwidth}
   \centering \includegraphics[width=4.6cm,height=3.5cm]{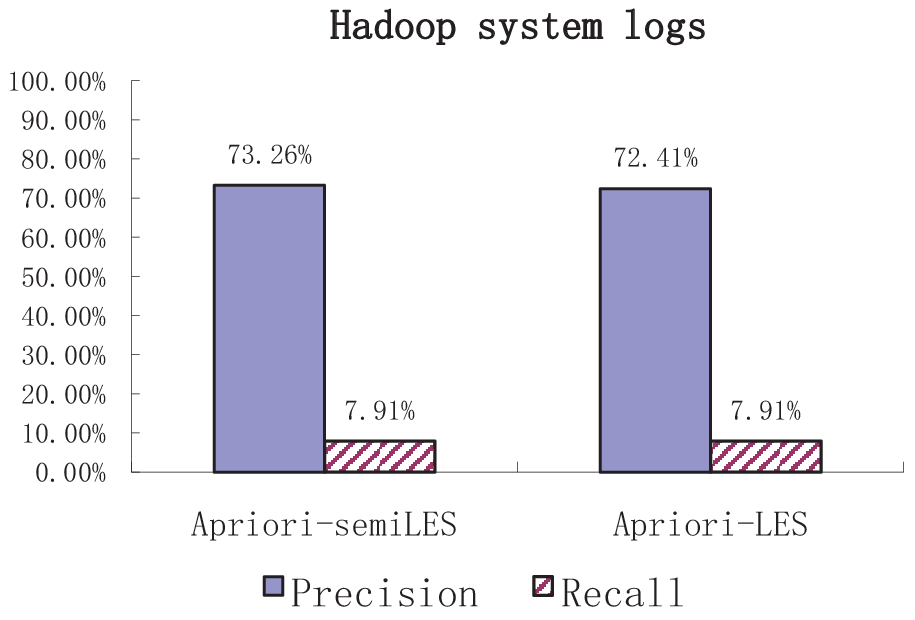}
   \end{minipage}%
   \begin{minipage}[c]{0.25\textwidth}
   \centering \includegraphics[width=4.6cm,height=3.5cm]{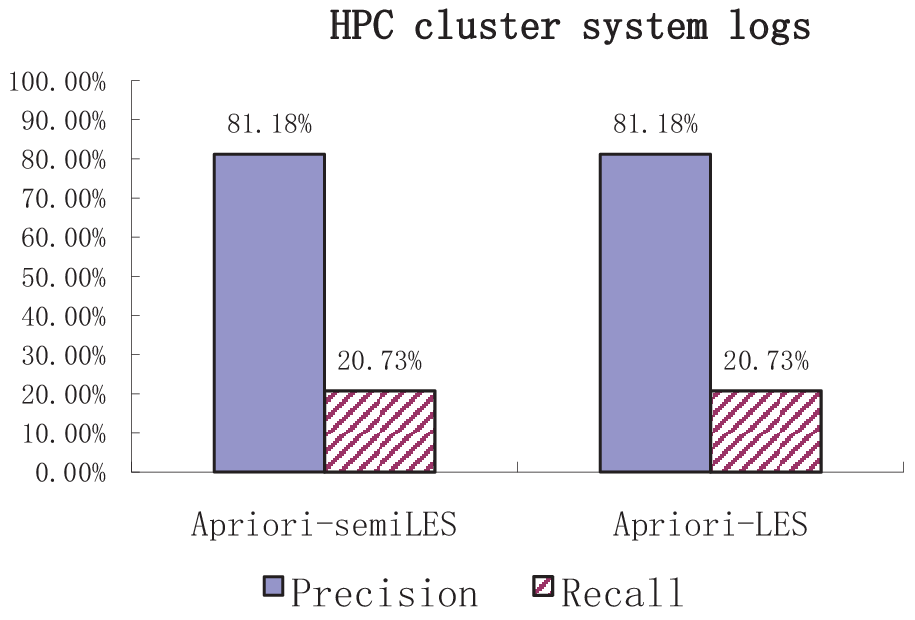}
   \end{minipage}
   \caption{The precision rate and recall rate of predicting failure events after removing non-failure events.}
   \label{fig:NF_prediction}
   \end{center}
   \end{figure}

TABLE~\ref{nonfailure-result} reports the filtered events at different stages. With respect to TABLE ~\ref{filtering-result}, TABLE~\ref{nonfailure-result} show after removing non-failure events, only a small fraction of events are reserved: 1,993 v.s. 132,650 entries (not removing non-failure events) for the HPC logs; 3,112 v.s. 26,538 entries (not removing non-failure events) for the Hadoop logs.    Then we predict failure events after removing non-failure events, and the precision rates and recall rates of predicting failure events are shown in Fig.~\ref{fig:NF_prediction}.

From Figs.\ref{fig:failure_prediction} and \ref{fig:NF_prediction}, we can observe two points.  First, after removing non-failure events, the precision rates in predicting failures are lower than that without removing non-failure events.
The reason is that in both logs the numbers of the event rules that identify correlations between non-failure events
  and failure events are higher than that of the event rules that identify the correlations between different failure events as shown in TABLE~\ref{failure-rules} in Section \ref{rule_summary}.
Second, in predicting failures ( FAILURE and FATAL), the recall rates are low (especially for the Hadoop logs). This observation has two reasons. (a) Some events are independent, and they have not correlated events.
(b) Because of the setting of the sliding time window, the support count and the posterior count threshold values, some weak-correlated or long time-correlated
event rules may be discarded.
 			
\begin{table}[hbtp]
\renewcommand{\arraystretch}{1.3}
\centering
\begin{tabular}{|l|p{0.8cm}|p{1.05cm}|p{1.0cm}|p{1.05cm}|p{1.2cm}|}
  \hline
  \itshape logs& \itshape raw logs &\itshape Removing repeated events &\itshape Removing non-failure events  &\itshape Removing periodic events &\itshape Compression rate \\ \hline
  \itshape Hadoop& \itshape 977,858 &\itshape 375,369&\itshape 53,259 &\itshape 3,112&\itshape 99.68\%	\\ \hline
  \itshape HPC &\itshape 433,490 &\itshape 152,112 &\itshape  5,427 &\itshape 1,993 &\itshape 99.54\% \\ \hline
\end{tabular}
\caption{The results of filtered events after removing non-failure events. }
\label{nonfailure-result}
\end{table}


   With the same baseline configurations,  the average prediction time of two logs is shown in TABLE~\ref{average_time}, which indicates that system administrators or autonomic systems should have enough time to monitor and handle predicted events.


\begin{table}[hbp]
\begin{tabular}{|l|p{1.2cm}|p{1.2cm}|}
\hline
\itshape Average prediction time(minutes) &\itshape Hadoop logs& \itshape HPC cluster logs \\ \hline
\itshape event prediction&\itshape 42.78& \itshape 4.01 \\ \hline
\itshape failure prediction&\itshape 52.01 & \itshape 25.57 \\ \hline
\itshape failure prediction after removing non-failure&\itshape 52.01 & \itshape 25.57 \\ \hline
\end{tabular}
\caption{The average prediction time.}
\label{average_time}
\end{table}

\section{Related Work} \label{sec:related_work}
We summarize the related work from five perspectives: characterizing failure characteristics, log preprocessing, event correlation mining, anomaly (failure or performance bottleneck) prediction, and failure diagnosis.

\subsection {characterizing failure characteristics}

It has been long recognized that failure events are correlated, not independent. For example, in the year of 1992, Tang {\em et al.} \cite{Tang_TOC92} concluded that the impact of correlated failures on dependability is significant. The work in \cite{Sahoo_DSN04} has observed that there are strong spatial correlations between failure events and most of the failure events occur on a small fraction of the nodes. The work in \cite{Liang_DSN05} presents that some failure events such as network failure events and application I/O failure events show more pronounced skewness in the spatial distribution. The work of \cite{Fu_ICS07} and \cite{Liang_DSN05} has found that failure events can propagate in the systems.

Some work uses statistical analysis approach to find simple temporal and spatial laws or models of system events \cite{Liang_DSN06} \cite{Fu_ICS07} \cite{Hacker_JPDC09} in large-scale cluster systems like Bluegene/L.  When the obtained knowledge is used in failure prediction, it may bring good precision rate and recall rate, but the prediction results are coarse and high level without the detail.

Daniel {\em et al.} \cite{Daniel} characterize the availability properties of cloud storage systems based on an extensive one year study of Google¡¯s main storage infrastructure and present statistical models that enable further insight into the impact of multiple design choices, such as data placement and replication strategies.

Edmund {\em et al.} \cite{Edmund} present the first large-scale analysis of hardware failure rates on a million consumer PCs. They found that many failures are neither transient nor independent. Instead, a large portion of hardware induced failures are recurrent: a machine that crashes from a fault in hardware is up to two orders of magnitude
more likely to crash a second time.

\subsection{Log Pre-processing}

Zheng {\em et al.} \cite{Zheng_DSN09} proposes a log pre-processing method, and adopt a causality-related filtering approach to combining correlated events for filtering through apriori association rule mining.

 In these research efforts, the concept of event cluster \cite{Liang_DSN05} is proposed to deal with multiple redundant records of fatal events at one location for event filtering; an apriori association rule mining \cite{Hellerstein_IBMSystemJournal02} is presented to identify the sets of fatal events co-occurring frequently and filter them together; an automated soft competitive learning neural-gas method \cite{Hacker_JPDC09} is used for cluster analysis to reduce dependent events.

 \subsection {Event Correlation Mining}

 In the data mining field, \cite{Yan_SDM03} \cite{Tzvetkov_ICDM03} concern about mining closed sequential patterns, \cite{Pei_KDD00} \cite{Mannila_DKD97} discusses the frequent pattern mining, \cite{Brin_DMKD98} \cite{Agrawal_SIGMODE93} \cite{Agrawal_VLDB94} focus on generalizing association rules to correlations.

Hellerstein {\em et al.} \cite {Hellerstein_IBMSystemJournal02} present efficient algorithms to mine three types of important
patterns from historical event data: event bursts, periodic patterns, and mutually dependent patterns, discuss a
framework for efficiently mining events that have multiple attributes, and finally build a tool---Event Correlation Constructor that validates and extends correlation knowledge.

Lou  \cite{ Lou_WASL092} propose an approach to mine inter-component dependencies from unstructured logs: parse each log message into keys and parameters; find dependent log key pairs belong to different components by leveraging co-occurrence analysis and parameter correspondence; use Bayesian decision theory to estimate the dependency direction of each dependent log key pair.

Mannila {\em et al.} \cite{Mannila_DKD97} give efficient algorithms for the discovery of all frequent episodes from a given class of episodes, and present detailed experimental results.

Though lots of previous efforts have proposed failures mining approaches for different purposes, for example event filtering \cite{Liang_DSN05}\cite{Zheng_DSN09}, event coalescing \cite{Hacker_JPDC09}, or failure prediction \cite{Sahoo_SIGKDD03}, little work proposes the event correlation mining system for large-scale cluster systems.

\subsection{Anomaly (Failure or Performance Bottleneck) Prediction}

\subsubsection{Performance bottleneck prediction}

Zhang {\em et al.} \cite{Zhang_DSN09} proposes a precise request tracing algorithm for multi-tier services of black boxes, which only uses application-independent knowledge and constructs a component activity graph abstraction to represent causal paths of requests and facilitate end-to-end performance debugging.

Gu {\em et al.} \cite{Gu_ICDE09} focus on predicting the bottleneck anomaly, the most common anomaly in data stream processing clusters. Their approach integrates naive Bayesian classification method, which captures the distinct symptoms of different bottlenecks caused by various reasons, and Markov models, which capture the changing patterns of different measurement metrics that are used as features by the Bayesian classifiers, to achieve the anomaly prediction goal.

Tan {\em et al.} \cite{Tan_PODC09} presents the context-aware anomaly prediction model
training algorithm to predict various system anomalies such as performance bottlenecks, resource hotspots, and service
level objective (SLO) violations. They first employ a clustering algorithm to discover different execution contexts in dynamic systems, and then train a set of prediction models, each of which is responsible for predicting anomalies under a specific context.

Shen {\em et al.} \cite{Shen_FAST05} propose a model-driven anomaly characterization approach and use it to discover operating system performance bugs when supporting disk I/O-intensive online servers.

\subsubsection{Failure prediction}

Gujrati {\em et al.} \cite{Gujrati_ICPP07} presents a meta-learning method based on statistical analysis and standard association rule algorithm. They not only obtain the statistical characteristics of failures,  but also generate association rules between nonfatal and fatal events for failure predictions.

Fu {\em et al.} \cite{Fu_ICS07}\cite{Fu_SRDS07} develops a spherical covariance model with an adjustable timescale parameter to quantify the temporal correlation and a stochastic model to describe spatial correlation. They cluster failure events based on their correlations and predict their future occurrences.

Fulp {\em et al.} \cite{ Fulp_WASL05} describes a spectrum-kernel Support Vector Machine (SVM) approach to predict failure events based on system log files. The approach described use a sliding window (sub-sequence) of messages to predict the
likelihood of failure.


\subsection{Failure Diagnosis}

Chen {\em et al.} \cite{Chen_SRDS10} presents an instance based approach to diagnosing failures in computing systems. Their method takes advantage of past experiences by storing historical failures in a database and developing a novel algorithm to efficiently retrieve failure signatures from the database.

Oliner {\em et al.} \cite{Oliner_DSN10} propose a method for identifying the sources of problems in complex production systems where, due to the prohibitive costs of instrumentation, the data available for analysis may be noisy or incomplete.

John {\em et al.} \cite{John_SRDS10} present a fault localization system called Spotlight that essentially uses two basic ideas. First, it compresses a multi-tier dependency graph into a bipartite graph with direct probabilistic edges between root causes and symptoms. Second, it runs a novel weighted greedy minimum set cover algorithm to provide fast inference.

Console logs rarely help operators detect problems in large-scale datacenter services, for they often consist of the voluminous intermixing of messages from many software components written by independent developers. Xu {\em et al.} \cite{Xu_SOSP09} \cite{Xu_SLAML10} propose a general methodology to mine this rich source of information to automatically detect system runtime problems.

Tucek {\em et al.} \cite{Tucek_SOSP07} propose a system Triage, that automatically performs onsite software failure diagnosis at the very moment of failure. It provides a detailed diagnosis report, including the failure nature, triggering conditions, related code and variables, the fault propagation chain, and potential fixes.

Tan {\em et al.} \cite{Tan_WASL08} propose SALSA---their approach to automated system-log analysis, which involves examining the logs to trace control-flow and data-flow execution in a distributed system, and derive state-machine-like views of the system¡¯s execution on each node. Based on the derived state machine views and statistics, they illustrate SALSA's value by developing visualization and failure-diagnosis techniques for three Hadoop workloads.

\section{ CONCLUSION AND FUTURE WORK} \label{sec:conclusion}

In this paper, we designed and implemented an event correlation mining system and an event prediction system. We presented a  simple metrics to measure correlations of events that may happen interleavedly.
On the basis of the measurement of correlations, we proposed two approaches to mining event correlations;
meanwhile, we  proposed an innovative abstraction---event correlation graphs (ECGs) to represent event correlations, and presented an ECGs-based algorithm for event prediction.
As two typical case studies, we used LogMaster to analyze and predict logs of a production Hadoop-based cloud computing system at Research Institution of China Mobile, and a production HPC cluster system at Los Alamos National Lab (LANL), respectively. For the first time, we compared the breakdown of events of different types and events rules in two typical cluster systems for Cloud and HPC, respectively.

In the new future, we will investigate two issues. a) How to use ECGs for fault diagnose in large-scale production cluster systems? b) How to combine causal path-based solutions \cite{Zhang_DSN09} with the log mining approach to diagnosis failure events and performance problems?



\appendices

\section {Choosing the threshold values of the periodic count and the periodic ratio in filtering periodic events}\label{appendix_preprocessing}


   \begin{figure}[htbp]
   \begin{center}
   \begin{minipage}[c]{0.25\textwidth}
   \centering
   \includegraphics[width=4.6cm,height=3.2cm]{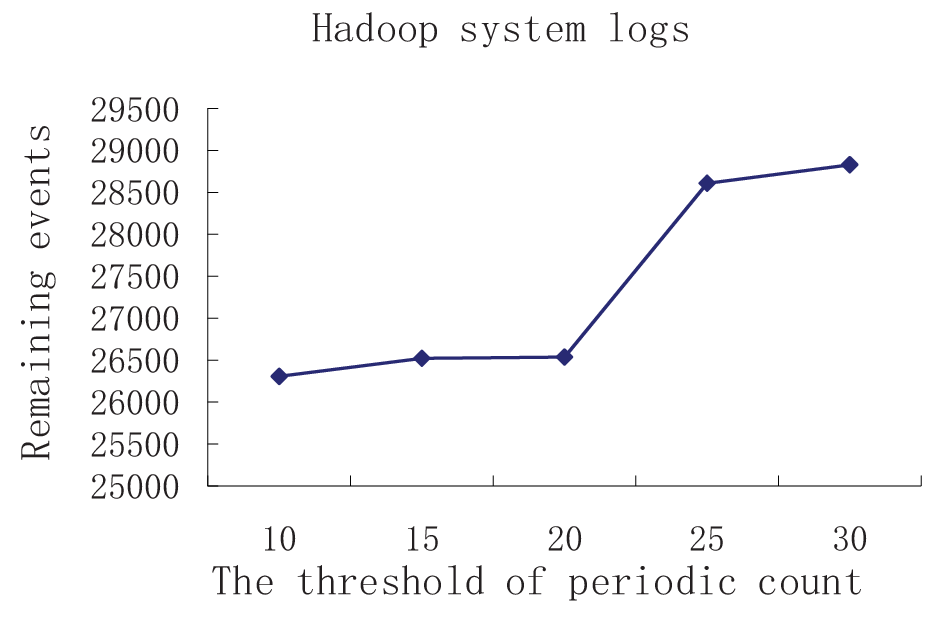}
   \includegraphics[width=4.6cm,height=3.2cm]{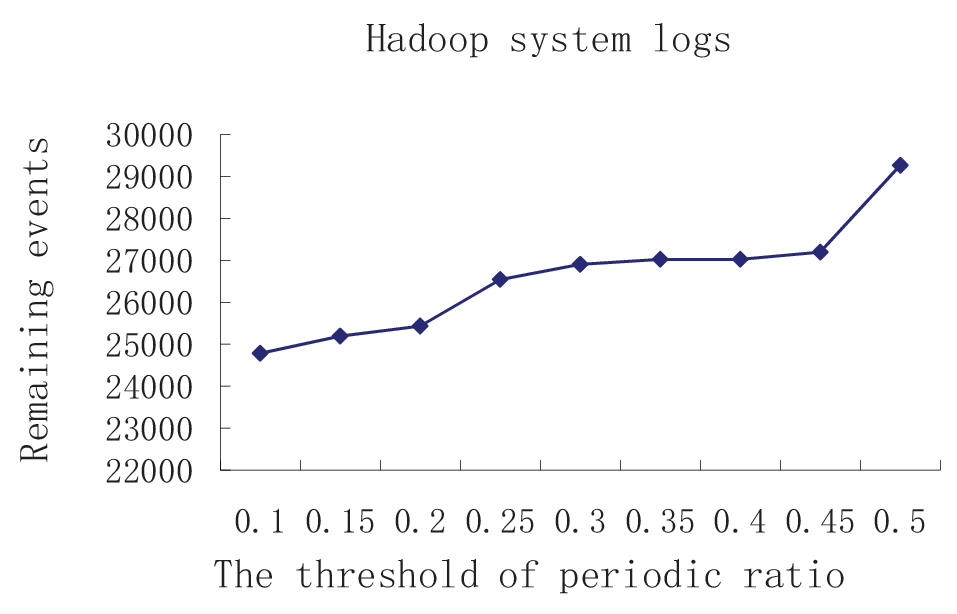}
   \end{minipage}%
   \begin{minipage}[c]{0.25\textwidth}
   \centering
   \includegraphics[width=4.6cm,height=3.2cm]{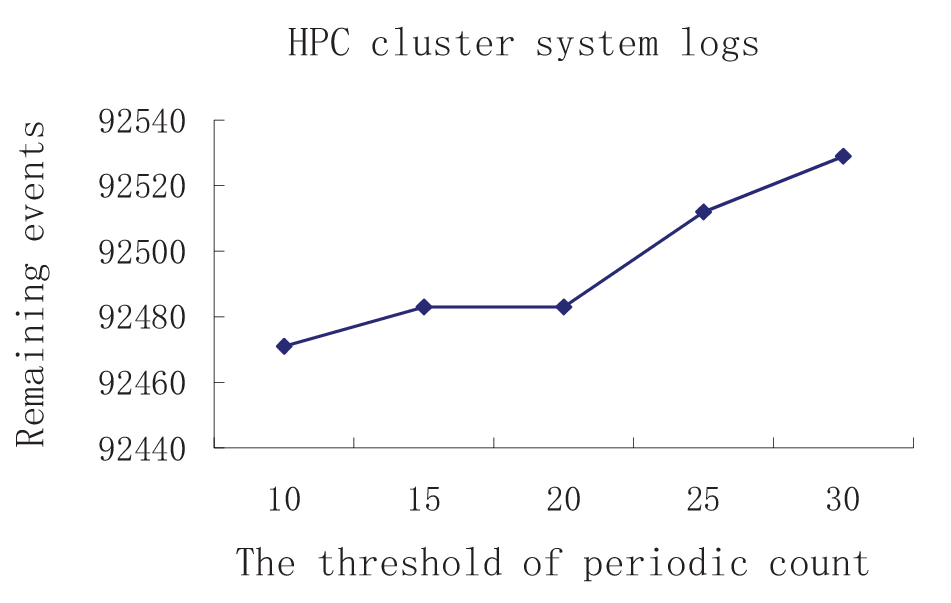}
   \includegraphics[width=4.6cm,height=3.2cm]{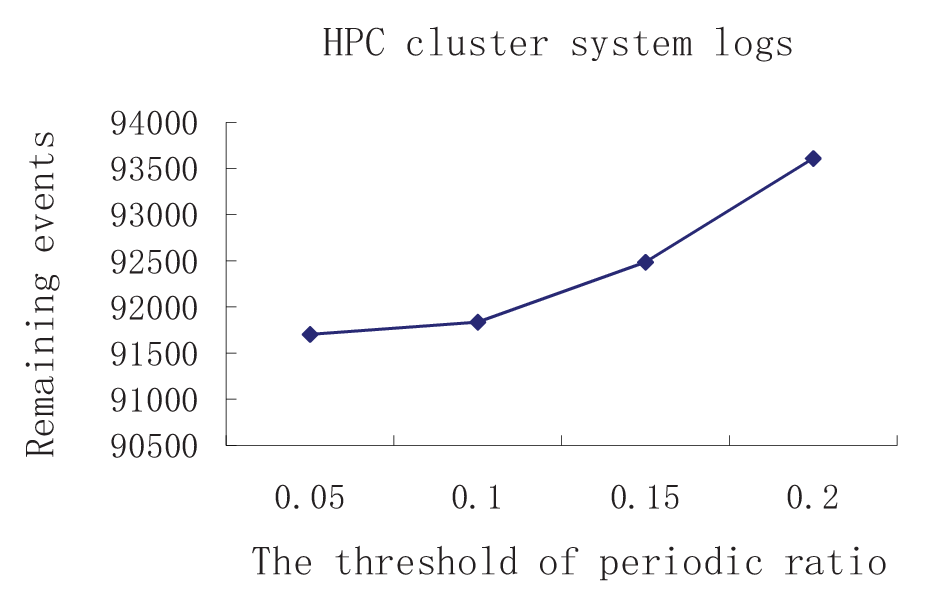}
   \end{minipage}
   \caption{ Relationships between remaining events and two thresholds¡ªperiodic count and  periodic ratio.}
   \label{fig:period}
   \end{center}
   \end{figure}

As shown in Fig. \ref{fig:period}, when the threshold of the periodic count is less than 20, the remaining numbers of Hadoop system logs and HPC cluster system logs change little.  When the threshold of the periodic count is above 20, the remaining events
change significantly, so we set the threshold of the periodic count as 20.

When the threshold of the periodic ratio of Hadoop system logs is less than 0.2 and the threshold of periodic ratio
of HPC cluster system logs is less than 0.1, the remaining events change little. So we set
the threshold of the periodic ratio of Hadoop system logs to 0.2 and the threshold of periodic ratio of
HPC cluster system logs to 0.1, respectively.

\begin{table}[hbtp]
\renewcommand{\arraystretch}{1.3}
\centering
\begin{tabular}{|l|p{2.7cm}|p{2.7cm}|}
  \hline
  \itshape logs & \itshape the threshold of periodic count &\itshape the threshold of periodic ratio \\ \hline
  \itshape Hadoop & \itshape 20  &\itshape 0.2  \\ \hline
  \itshape HPC cluster & \itshape 20 &\itshape 0.1 \\ \hline
\end{tabular}
\caption{Threshold selection of two logs}
\label{Threshold-selection}
\end{table}
	 	

\section {Parameters effects in event correlation mining } \label{appendix_parameters}

\begin{figure*}[!htbp]
   \begin{center}
   \begin{minipage}[c]{0.3\textwidth}
   \centering
    \includegraphics[width=6cm,height=4cm]{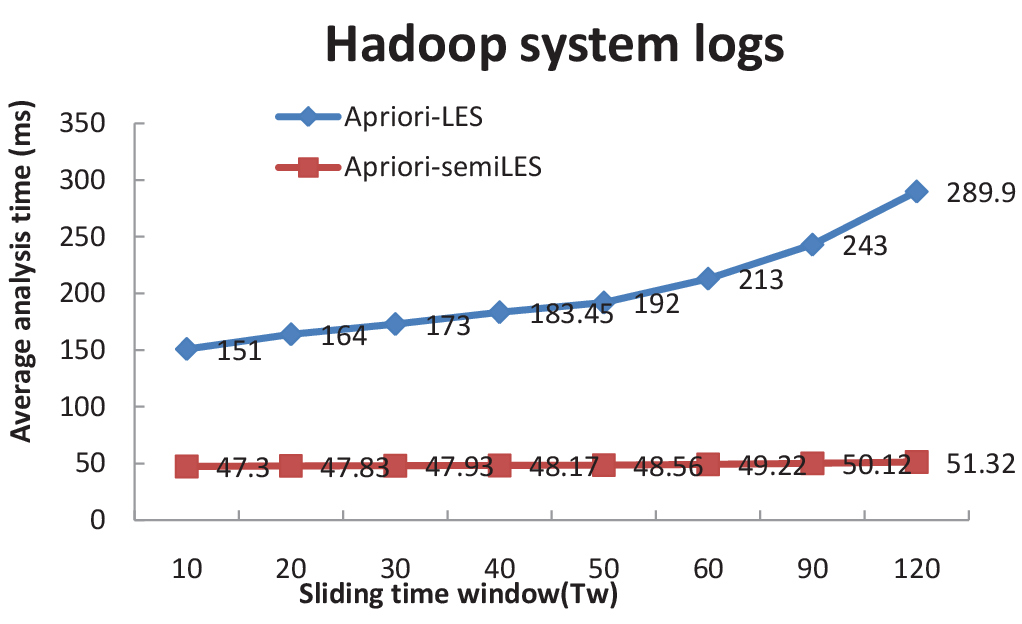}
    \includegraphics[width=6cm,height=4cm]{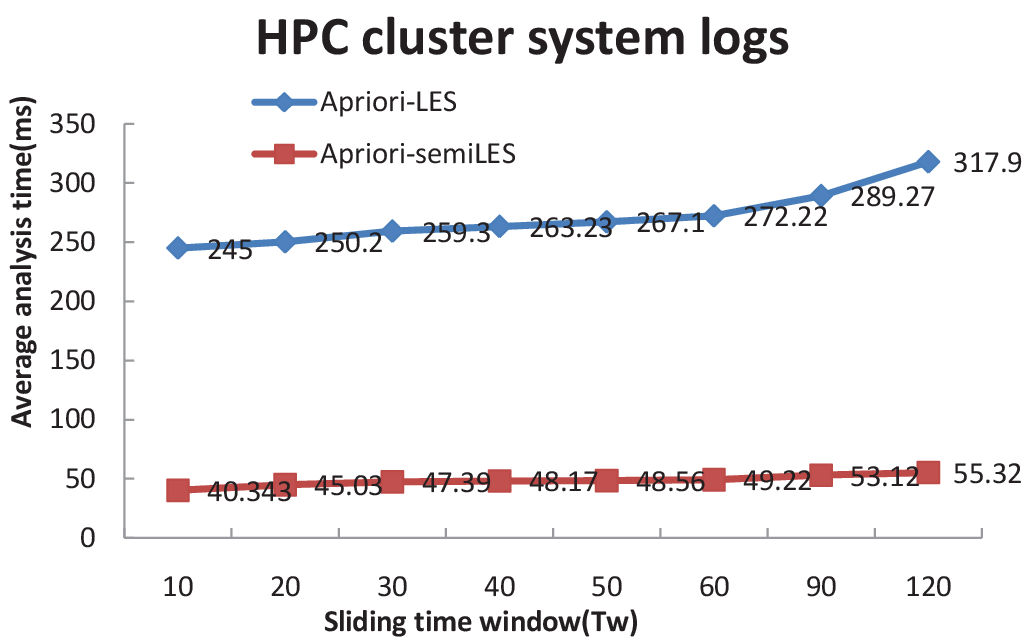}
   \end{minipage}%
   \begin{minipage}[c]{0.3\textwidth}
   \centering
   \includegraphics[width=6cm,height=4cm]{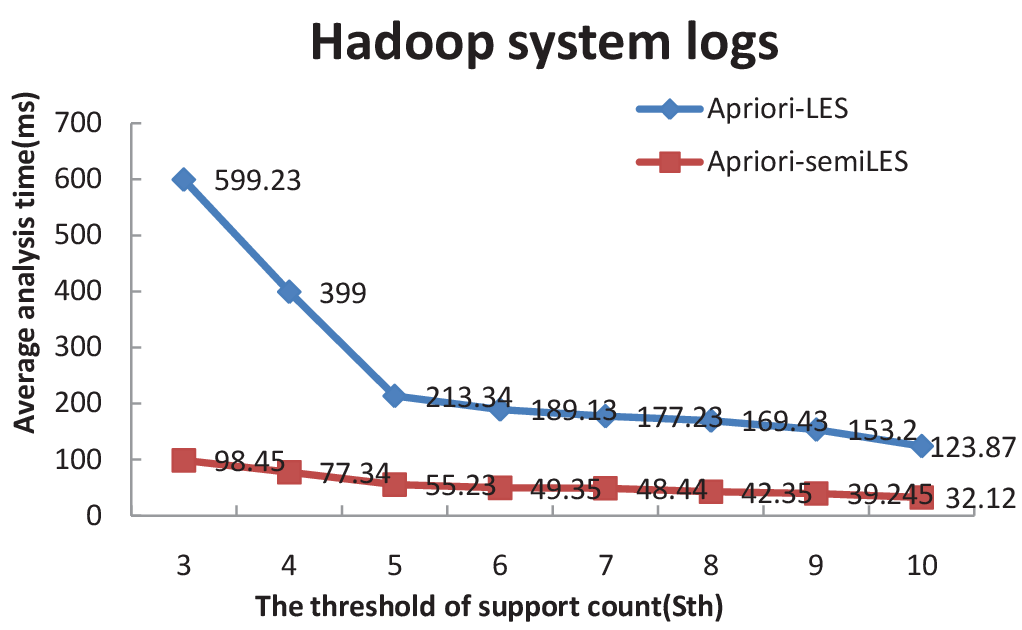}
     \includegraphics[width=6cm,height=4cm]{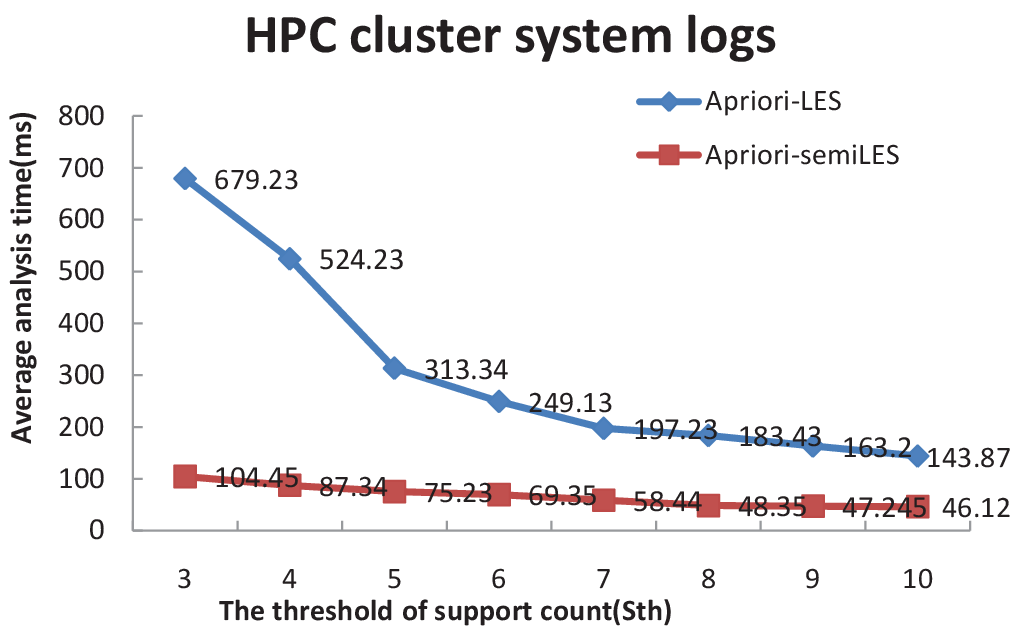}
   \end{minipage}
    \begin{minipage}[c]{0.3\textwidth}
   \centering
   \includegraphics[width=6cm,height=4cm]{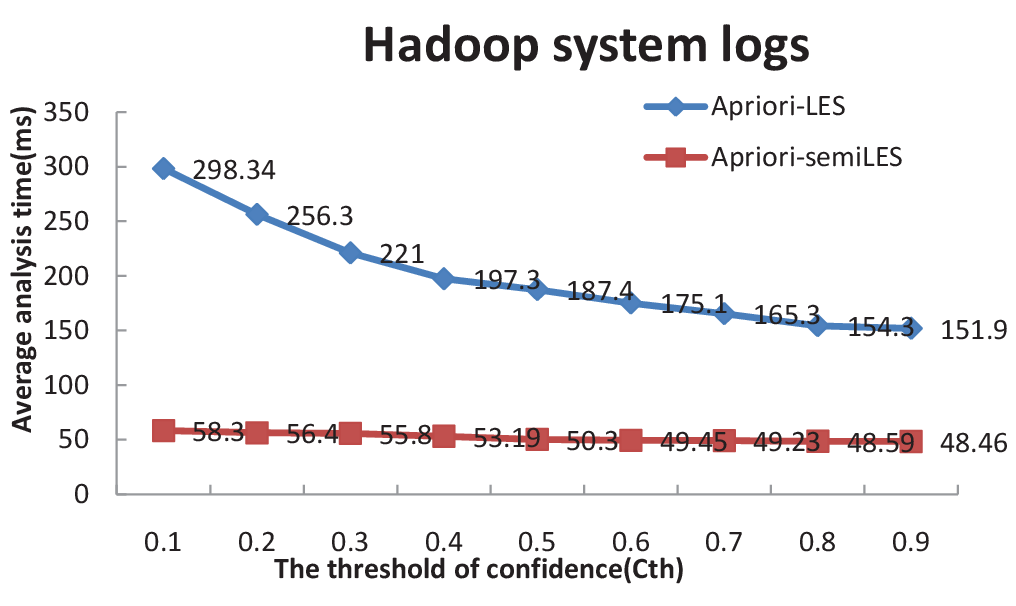}
   \includegraphics[width=6cm,height=4cm]{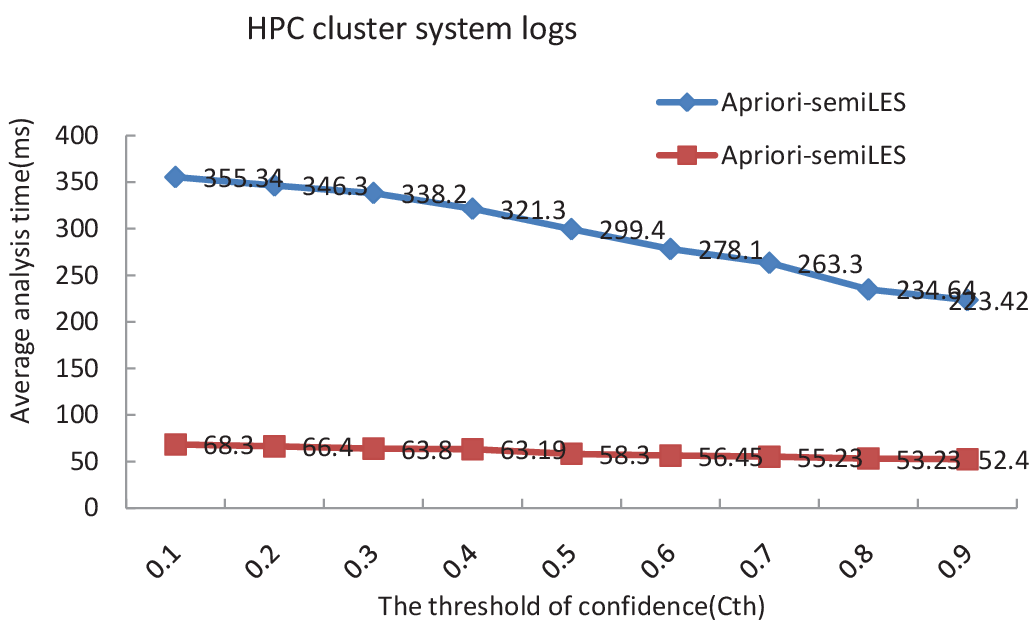}
   \end{minipage}
   \caption{\label{fig:effect_time} The effect of parameters(\emph{Tw}, \emph{Sth} and \emph{Cth}) on the average analysis time per event.}
   \end{center}
   \end{figure*}

\begin{figure*}[!htbp]
   \begin{center}
   \begin{minipage}[c]{0.3\textwidth}
   \centering
   \includegraphics[width=6cm,height=4cm]{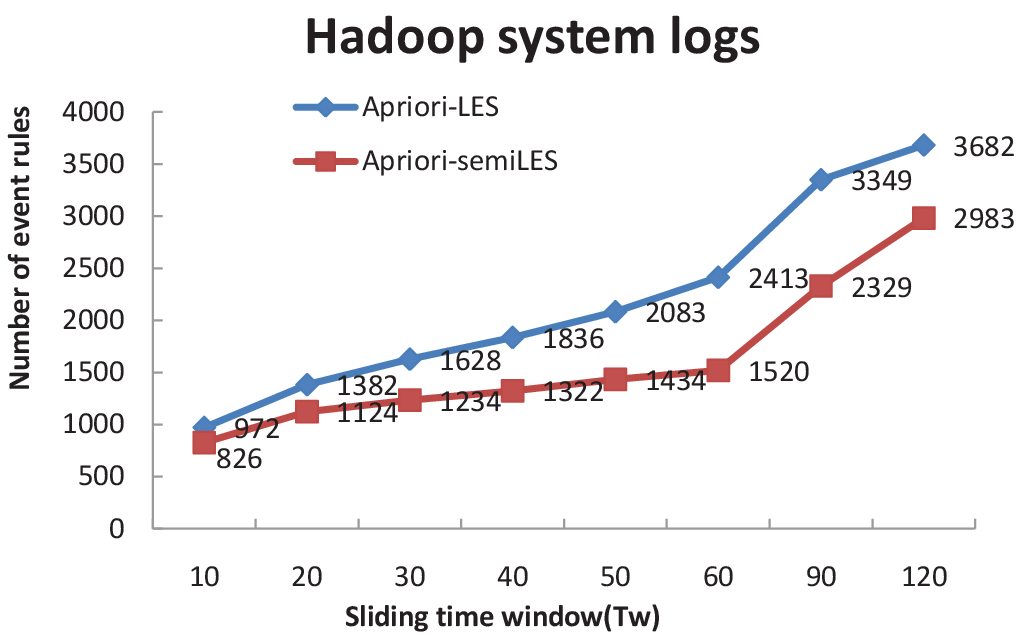}
   \includegraphics[width=6cm,height=4cm]{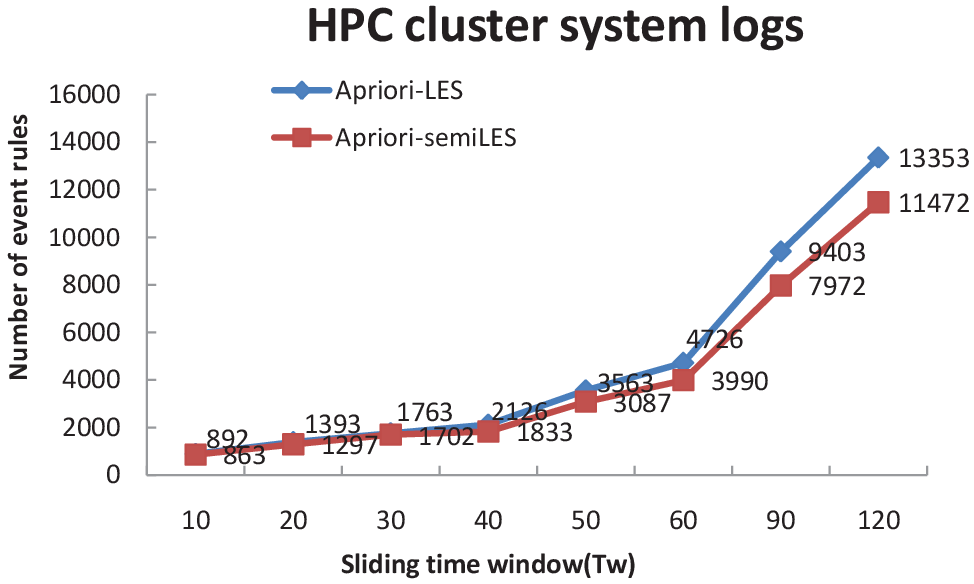}
   \end{minipage}%
   \begin{minipage}[c]{0.3\textwidth}
   \centering
    \includegraphics[width=6cm,height=4cm]{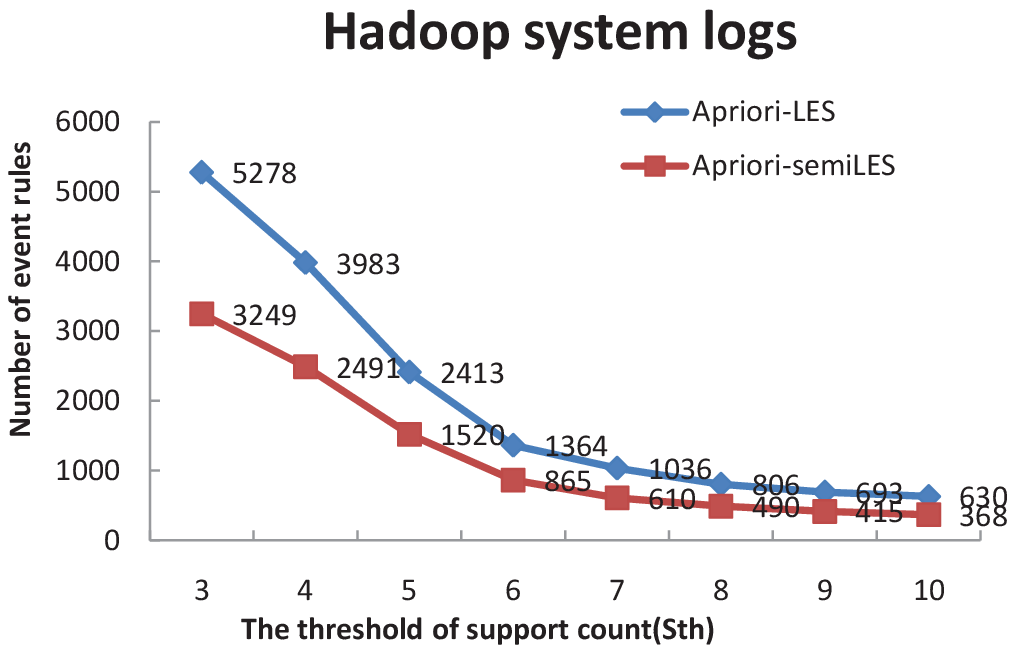}
   \includegraphics[width=6cm,height=4cm]{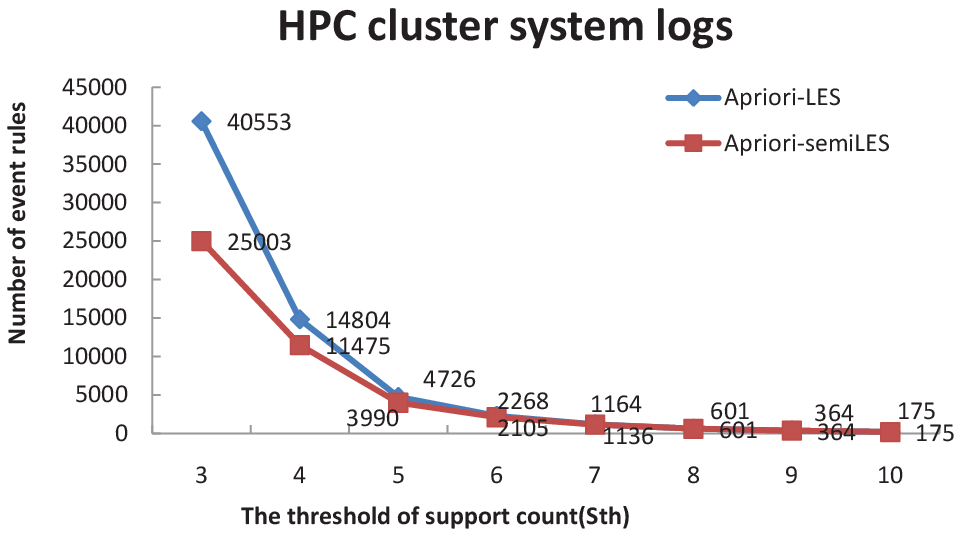}
   \end{minipage}
   \begin{minipage}[c]{0.3\textwidth}
   \centering
    \includegraphics[width=6cm,height=4cm]{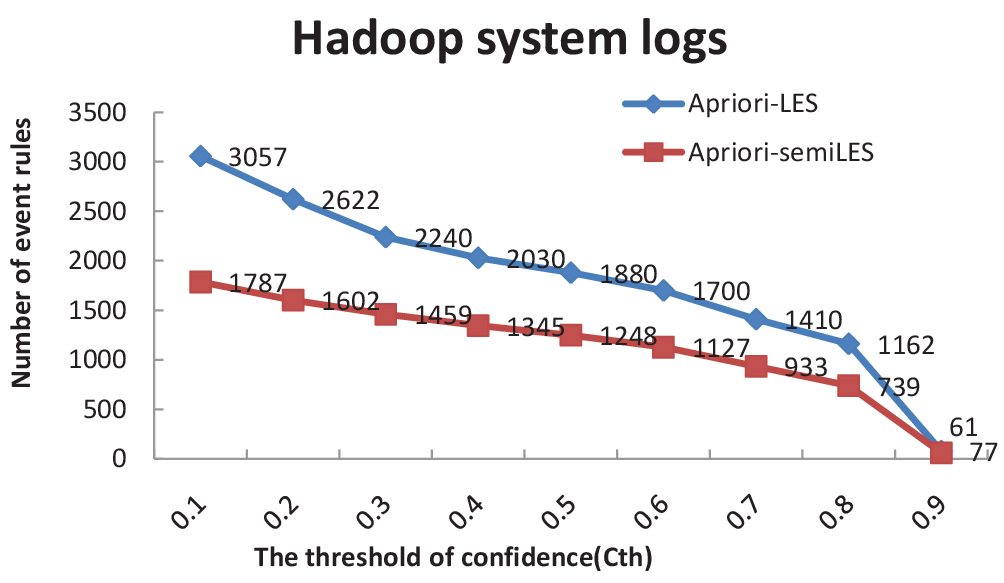}
   \includegraphics[width=6cm,height=4cm]{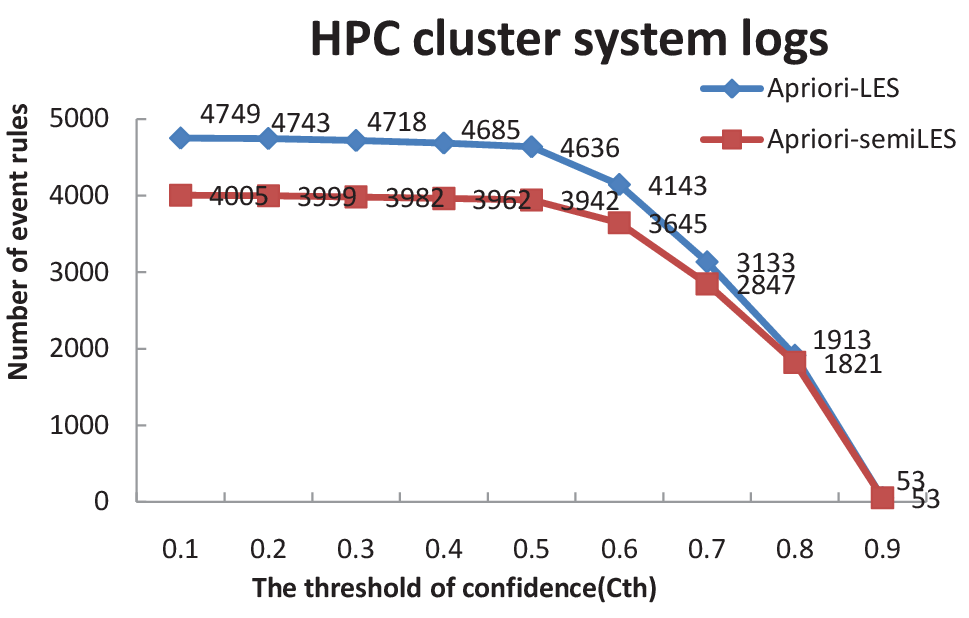}
   \end{minipage}
   \caption{\label{fig:effect_number} The effect of parameters(\emph{Tw}, \emph{Sth} and \emph{Cth}) on the number
   of event rules.}
   \end{center}
   \end{figure*}

The effects of parameters($Tw$, $Sth$ and $Cth$) on the average analysis time per event and the number
   of event rules are shown in Fig.~\ref{fig:effect_time} and Fig.~\ref{fig:effect_number}, respectively.

\section {Parameters effects in event predictions} \label{appendix_prediction_parameters}

\begin{figure*}[!htbp]
   \begin{center}
   \begin{minipage}[c]{0.3\textwidth}
   \centering
    \includegraphics[width=6cm,height=4cm]{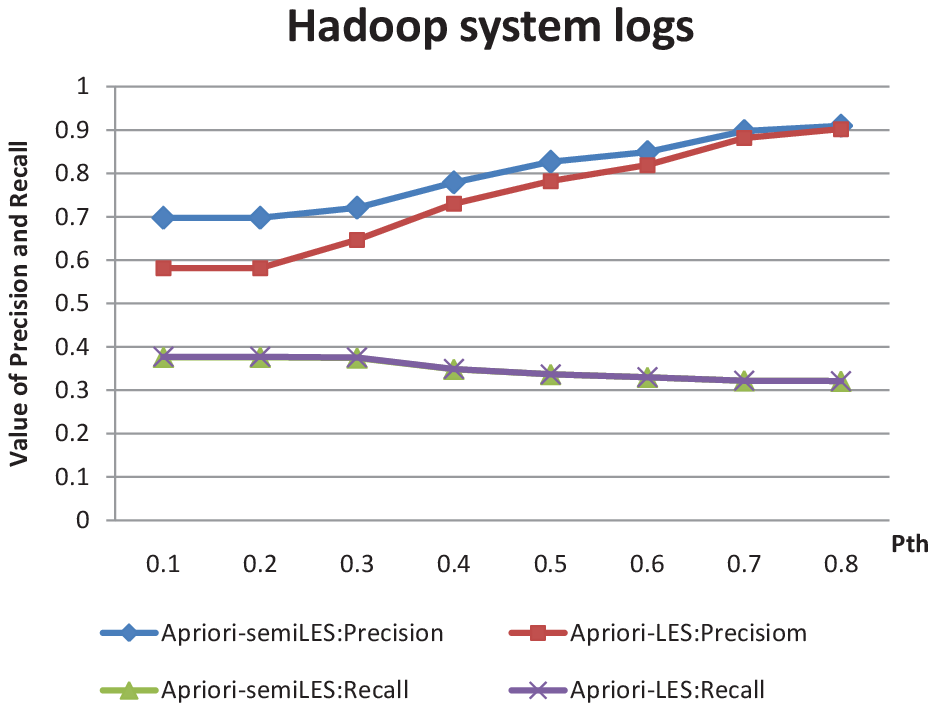}
    \includegraphics[width=6cm,height=4cm]{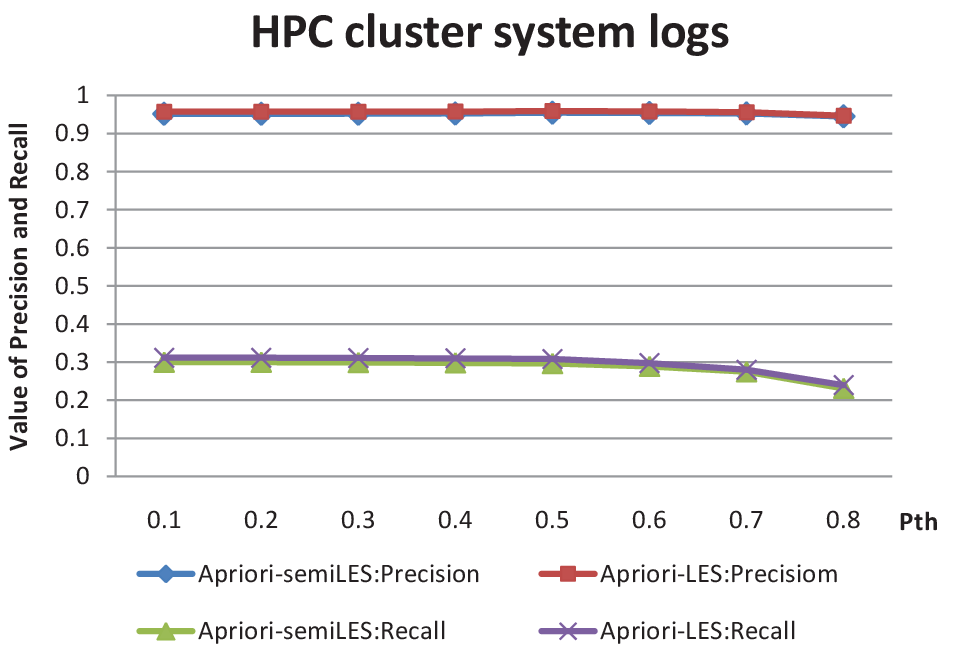}
   \end{minipage}%
   \begin{minipage}[c]{0.3\textwidth}
   \centering
   \includegraphics[width=6cm,height=4cm]{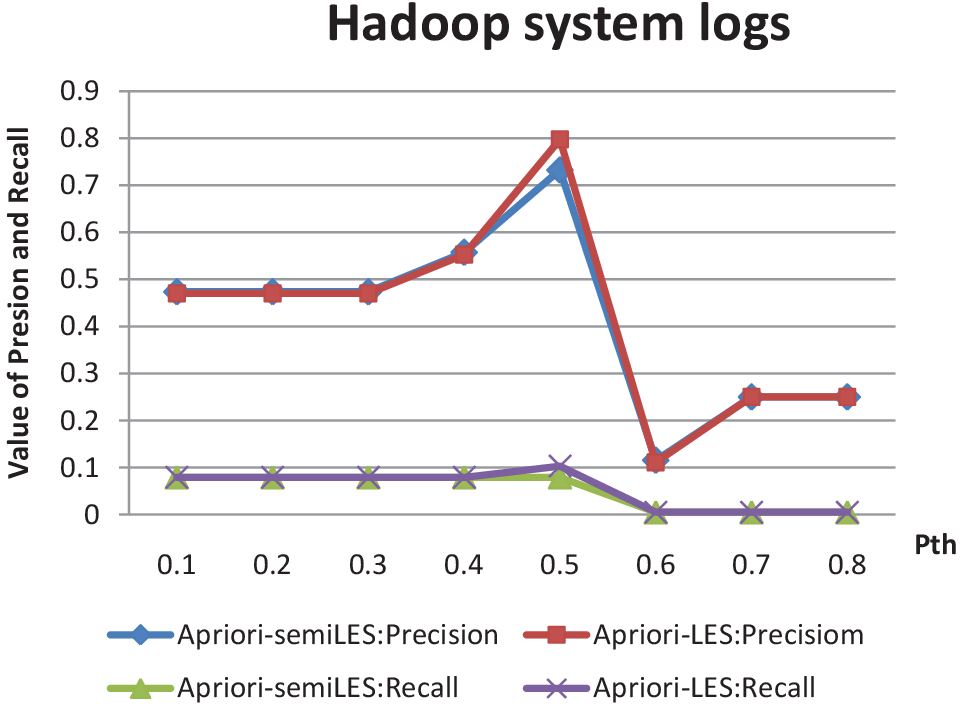}
     \includegraphics[width=6cm,height=4cm]{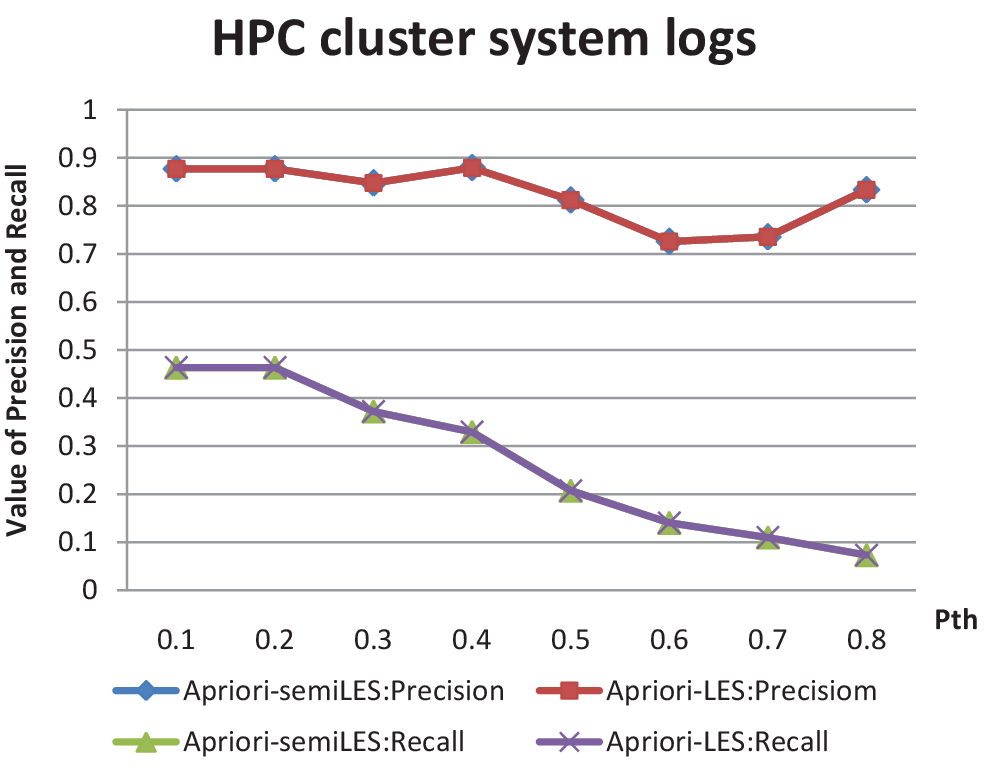}
   \end{minipage}
    \begin{minipage}[c]{0.3\textwidth}
   \centering
   \includegraphics[width=6cm,height=4cm]{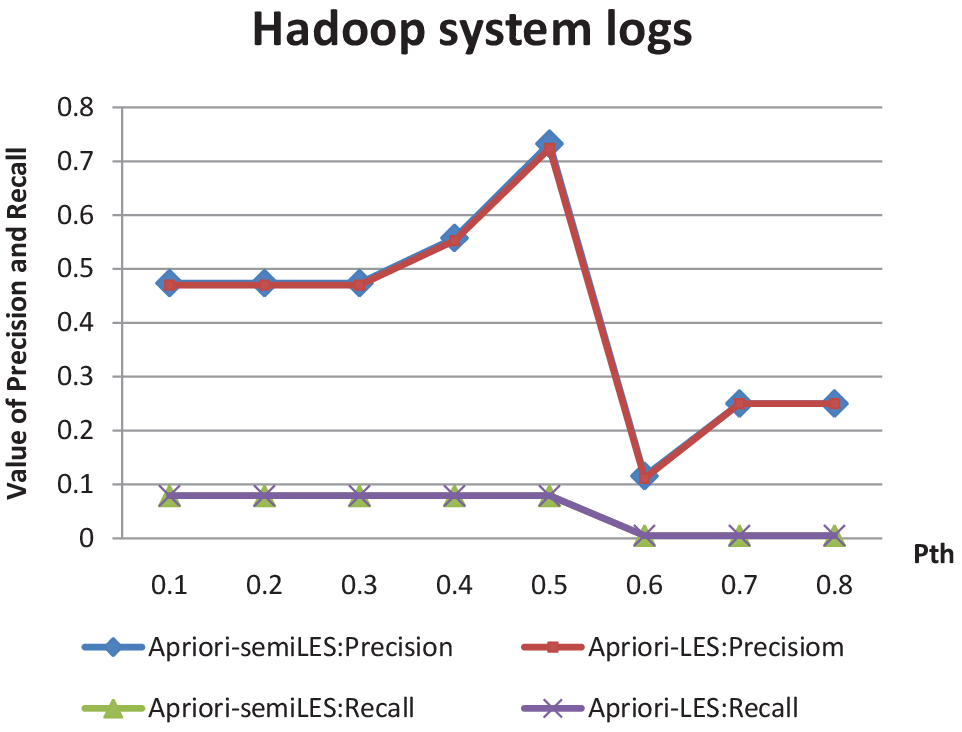}
   \includegraphics[width=6cm,height=4cm]{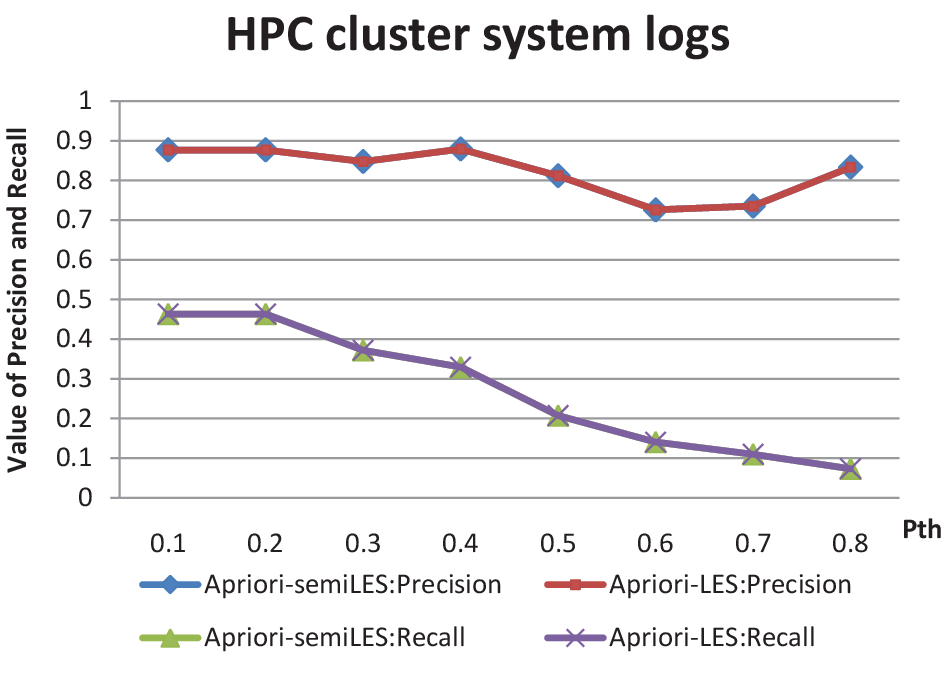}
   \end{minipage}
   \caption{\label{fig:effect_Pth} The effects of the parameter \emph{Pth} on the precision rate and the recall rate of the Hadoop system logs and the HPC cluster system logs in predicting all events on the basis of both failure and non-failure events, predicting only failure events on the basis of both failure and non-failure events, and predicting failure events after removing non-failure events (from left to right).}
   \end{center}
   \end{figure*}

\begin{figure*}[!htbp]
   \begin{center}
   \begin{minipage}[c]{0.3\textwidth}
   \centering
    \includegraphics[width=6cm,height=4cm]{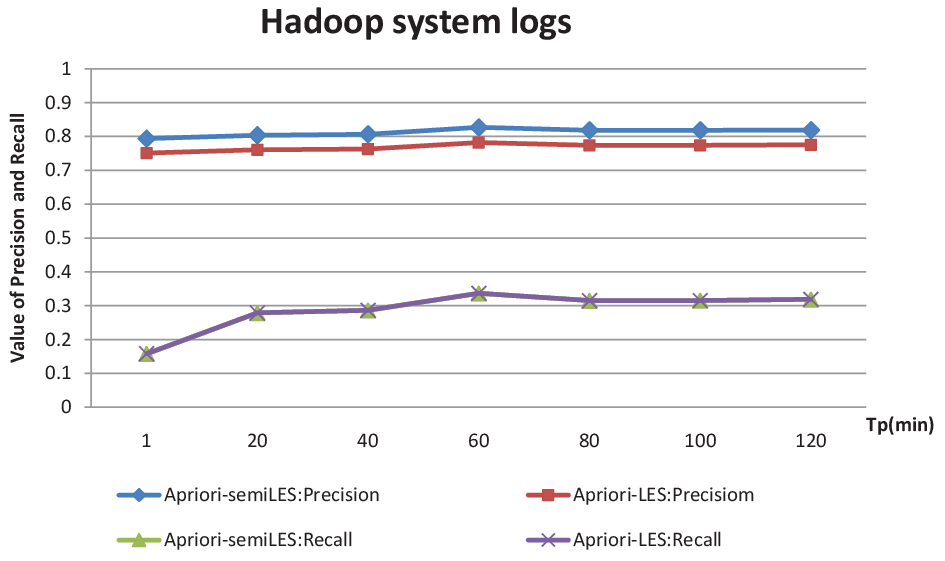}
    \includegraphics[width=6cm,height=4cm]{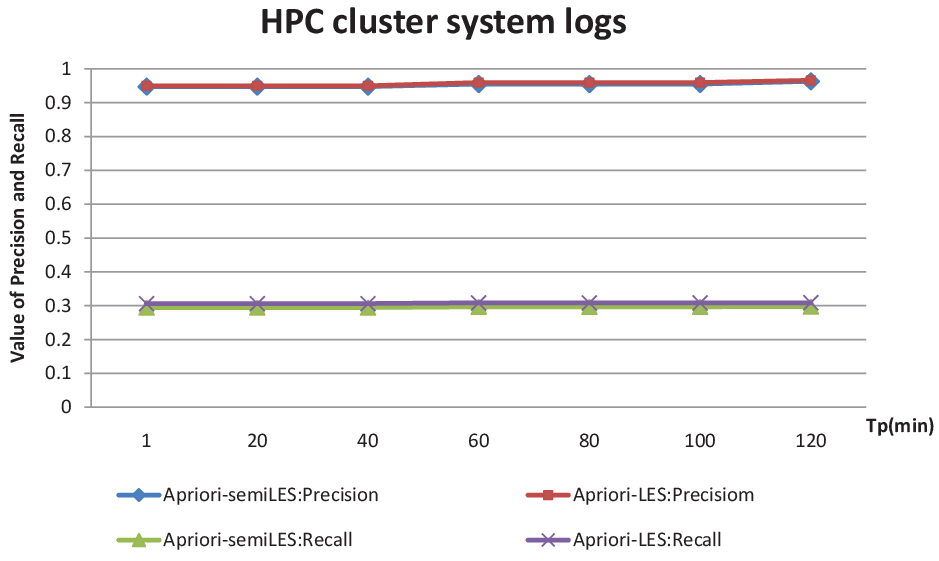}
   \end{minipage}%
   \begin{minipage}[c]{0.3\textwidth}
   \centering
   \includegraphics[width=6cm,height=4cm]{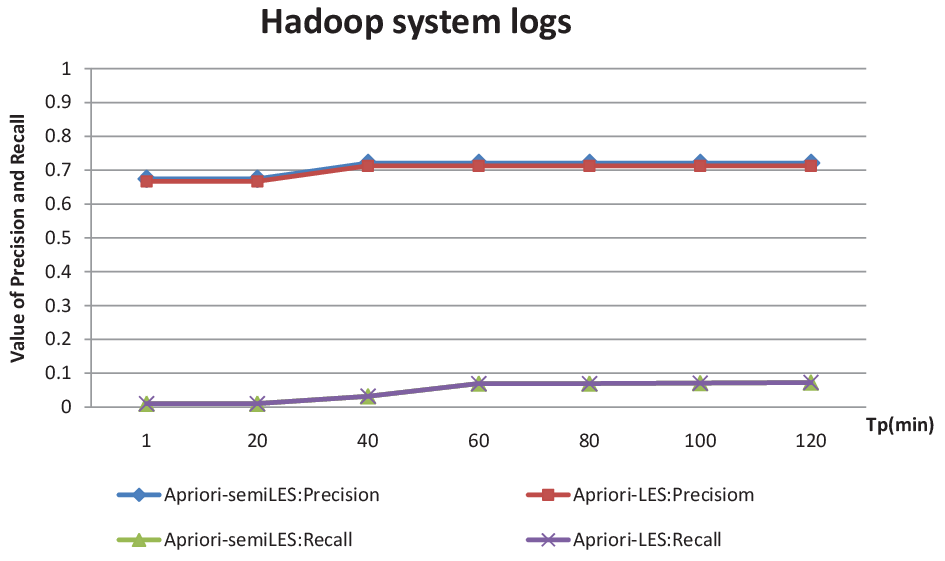}
     \includegraphics[width=6cm,height=4cm]{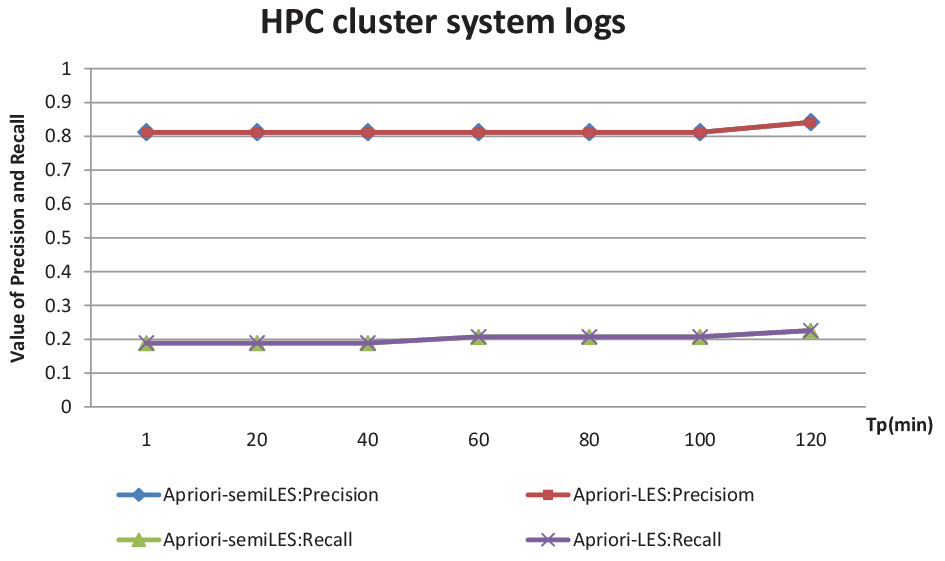}
   \end{minipage}
    \begin{minipage}[c]{0.3\textwidth}
   \centering
   \includegraphics[width=6cm,height=4cm]{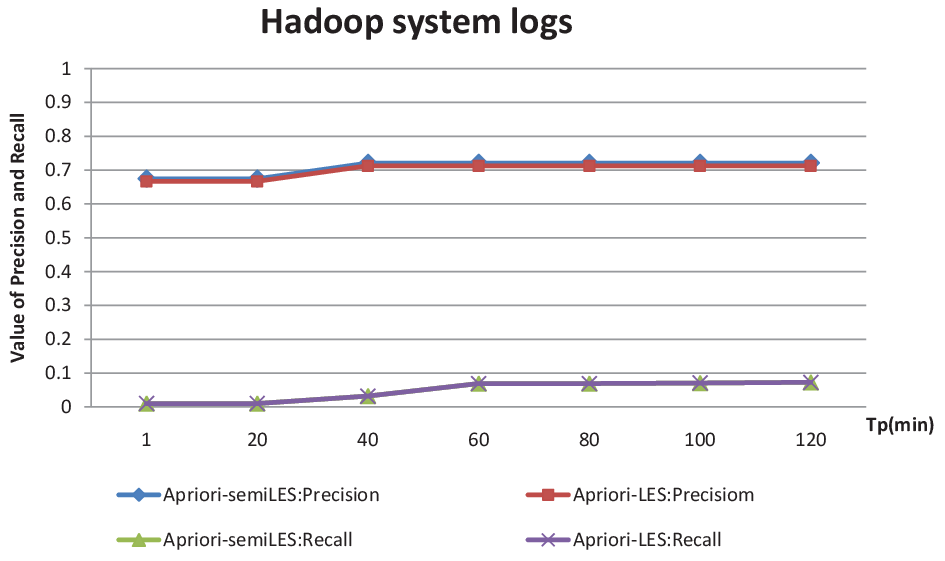}
   \includegraphics[width=6cm,height=4cm]{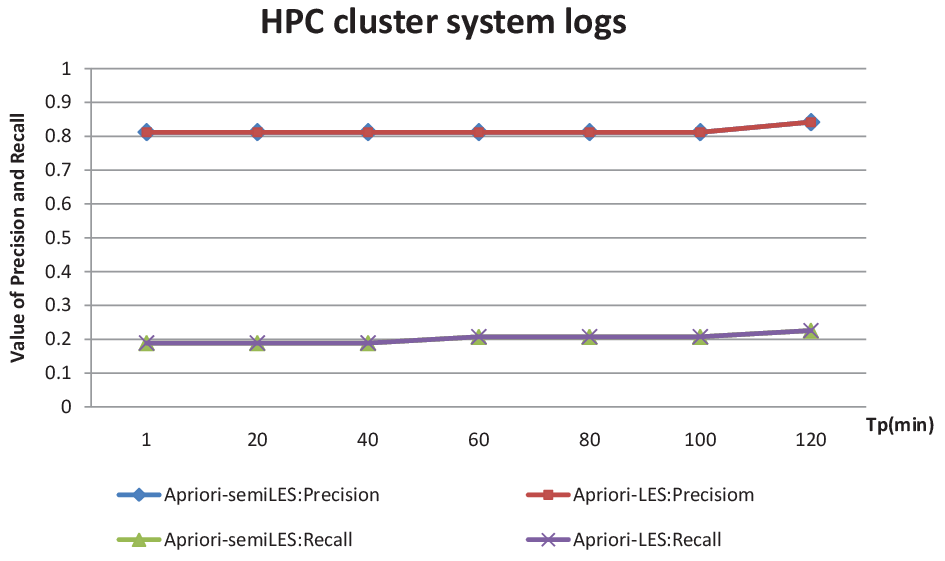}
   \end{minipage}
   \caption{\label{fig:effect_Tp} The effects of the parameter \emph{Tp} on the precision rate and recall rate of the Hadoop system logs and the HPC cluster system logs in predicting all events on the basis of both failure and non-failure events, predicting only failure events on the basis of both failure and non-failure events, and predicting failure events after removing non-failure events (from left to right).}
   \end{center}
   \end{figure*}

The effects of parameters($Pth$ and $Tp$) on the precision rates and recall rates are shown in Fig.~\ref{fig:effect_Pth}, Fig.~\ref{fig:effect_Tp}, respectively.

\end{document}